\documentclass[preprint,aps,prd,nofootinbib,superscriptaddress,eqsecnum,tightenlines,floatfix]{revtex4-2}
\usepackage[T1]{fontenc}
\usepackage{lmodern}
\usepackage{graphicx}
\graphicspath{{./modified_figures_v2/}}
\usepackage[caption=false]{subfig} 
\usepackage{amsmath,amssymb}
\usepackage{mathtools}
\usepackage{bm}

\usepackage{hyperref}
\usepackage{orcidlink}
\usepackage{zref-clever}
\usepackage{here}

\usepackage{xcolor}
\usepackage{tikz}
\usetikzlibrary{shapes.geometric,shapes.misc,arrows.meta,positioning,fit,backgrounds,shadows.blur}
\usepackage{booktabs,parskip,titlesec,fancyhdr,etoolbox,tabularx,array,graphicx}

\definecolor{PRDblue}   {HTML}{1A56A4}
\definecolor{PRDgray}   {HTML}{475569}
\definecolor{PRDlight}  {HTML}{F1F5F9}
\definecolor{PRDborder} {HTML}{CBD5E1}
\definecolor{StageA}    {HTML}{0F4C81}
\definecolor{StageB}    {HTML}{1A6B5C}
\definecolor{StageC}    {HTML}{7B2D8B}
\definecolor{NodeBg}    {HTML}{FFFFFF}
\definecolor{NodeBrd}   {HTML}{1A56A4}
\definecolor{DiamBg}    {HTML}{EFF6FF}
\definecolor{DiamBrd}   {HTML}{0F4C81}
\definecolor{ParaBg}    {HTML}{FAF5FF}
\definecolor{ParaBrd}   {HTML}{7B2D8B}
\definecolor{ArrowClr}  {HTML}{475569}

\tikzset{
  proc/.style={rectangle,draw=NodeBrd,line width=1pt,fill=NodeBg,rounded corners=3pt,
    text width=2.9cm,align=center,font=\footnotesize\sffamily,text=PRDgray,inner sep=6pt},
  diam/.style={diamond,draw=DiamBrd,line width=1.2pt,fill=DiamBg,aspect=1.15,
    text width=2.5cm,align=center,font=\footnotesize\sffamily,text=PRDgray,inner sep=1pt},
  para/.style={trapezium,trapezium left angle=80,trapezium right angle=100,
    draw=ParaBrd,line width=1pt,fill=ParaBg,text width=2.8cm,align=center,
    font=\footnotesize\sffamily,text=PRDgray,inner sep=6pt},
  stage/.style={rectangle,draw=PRDborder,line width=0.7pt,fill=PRDlight,rounded corners=5pt,inner sep=10pt},
  arr/.style={->,>=Stealth,line width=1.3pt,color=ArrowClr,shorten >=2pt,shorten <=2pt}
}

\zcsetup{lang=english,cap=true,abbrev=true,nameinlink=tsingle}
\zcLanguageSetup{english}{
  namesep   = {\nobreak},
  rangesep  = {\textendash},
}
\zcRefTypeSetup{section}{
  Name-sg=Section,  Name-pl=Sections,
  Name-sg-ab=Sec.~,  Name-pl-ab=Secs.~,
}
\zcRefTypeSetup{figure}{
  Name-sg=Figure,   Name-pl=Figures,
  Name-sg-ab=Fig.~,  Name-pl-ab=Figs.~,
}
\zcRefTypeSetup{table}{
  Name-sg=Table,    Name-pl=Tables,
  Name-sg-ab=Table~, Name-pl-ab=Tables~,
}
\zcRefTypeSetup{equation}{
  Name-sg=Equation, Name-pl=Equations,
  Name-sg-ab=Eq.,   Name-pl-ab=Eqs.,
  refbounds={,(,),}
}
\zcRefTypeSetup{appendix}{
  Name-sg=Appendix,    Name-pl=Appendices,
  Name-sg-ab=Appendix~, Name-pl-ab=Appendices~,
}

\newcommand{\ORCID}[1]{%
  {\,\orcidlink{#1}}%
}

\newcounter{num}

\begin{document}

\title{Testing $\Lambda$CDM with ANN-Reconstructed Expansion History from Cosmic Chronometers}

\author{Yuki Hashimoto\ORCID{0000-0001-9624-2503}}
\email{s2471002@ipc.fukushima-u.ac.jp}
\affiliation{Faculty of Symbiotic Systems Science, Fukushima University, Fukushima 960-1296, Japan.}

\author{Kazuharu Bamba\ORCID{0000-0001-9720-8817}}
\email{bamba@sss.fukushima-u.ac.jp}
\affiliation{Faculty of Symbiotic Systems Science, Fukushima University, Fukushima 960-1296, Japan.}

\author{Sanjay Mandal\ORCID{0000-0003-2570-2335}}
\email{sanjaymandal960@gmail.com}
\affiliation{Faculty of Symbiotic Systems Science, Fukushima University, Fukushima 960-1296, Japan.}

\begin{abstract}
In modern cosmology, the rapid growth of high-precision observational data, along with significant theoretical advances, has intensified the challenge of identifying a robust, model-independent framework to probe the expansion history of the Universe. In this work, we propose a novel artificial neural network (ANN)-based framework for the non-parametric reconstruction of the late-time cosmic expansion. The framework is trained and validated through a three-stage screening pipeline prior to its application to real observational data. As a demonstration of its effectiveness, we reconstruct the Hubble parameter $H(z)$ using the latest cosmic chronometer measurements. Our results show that the reconstructed expansion history aligns with the predictions of the $\Lambda$CDM model within observational uncertainties, thereby supporting the robustness and reliability of the proposed approach.
\end{abstract}
\maketitle
\newpage
%
\section{\label{sec:intro}Introduction}
The standard $\Lambda$CDM model provides a remarkably successful effective description of a wide range of cosmological observations, including the Planck 2018 results~\cite{Planck:2020Params}, dark-energy reviews~\cite{Frieman:2008DEreview,Padmanabhan:2003LambdaReview}, and recent Baryon Acoustic Oscillation (BAO) constraints from DESI~\cite{DESI:2025fxa,DESI:2025BAO}. However, the late-time expansion history remains a key area in which to test that description with as few unnecessary assumptions as possible. In particular, the continuing discussion about the Hubble tension~\cite{Verde:2019Tensions,DiValentino:2021Realm,Perivolaropoulos:2021jda,Schoneberg:2021qvd,Vagnozzi:2023nrq,Riess:2022SH0ES,Freedman:2024CCHP} has strengthened the case for reconstruction strategies that do not rely exclusively on a low-dimensional parametric approach for $H(z)$. Such strategies are useful not only for testing the consistency of $\Lambda$CDM~\cite{Lei:2025zjx}, but also for providing a controlled interface between observational data and later model selection.
The Cosmic Chronometers (CC)~\cite{JimenezLoeb:2002RelativeAges,Simon:2005CC,Stern:2010CC,Moresco:2012CC,Moresco:2016CC6,Ratsimbazafy:2017CC,Borghi:2022CC,Jiao:2023LEGACC,Moresco:2020CCcov,Moresco:2022phi,Wang:2026lno} are especially suitable for a baseline reconstruction study because they provide direct estimates of the Hubble parameter through the differential-age relation. Unlike Type Ia supernovae distance-modulus data~\cite{Scolnic:2018Pantheon,Scolnic:2022PantheonPlus} or BAO measurements~\cite{Beutler:2011SixdF,Alam:2017DR12,Alam:2021eBOSS,DESI:2025BAO,DESI:2025fxa}, the CC data already live in the target space of this manuscript, namely the expansion rate $H(z)$ itself. This makes the CC compilation a natural starting point for an implementation-oriented study in which the main goal is to establish a reproducible Artificial Neural Networks (ANNs) baseline before adding more complicated multi-probe forward models.

Moreover, among non-parametric reconstruction methods, Gaussian Processes~\cite{Shafieloo:2007Smoothing,Seikel:2012GP,Zhang:2018GPDarkEnergy,Hwang:2022hla,Holsclaw:2010PRL,Holsclaw:2010PRD}, genetic algorithms~\cite{Nesseris:2012GA}, and neural-network-based reconstructions of the late-time observables~\cite{Wang:2020reconstruct,GomezVargas:2023nnrecon,Shah:2024LADDER} have all been used extensively in the late-time cosmology. ANNs provide a complementary route in which the regression function is learned from data rather than from an explicitly specified covariance kernel. In cosmology, this direction was established by the ReFANN framework~\cite{Wang:2020reconstruct}, where ANNs were used to reconstruct functions such as $H(z)$ and $D_L(z)$ in a non-parametric way and to infer cosmological parameters from the reconstructed functions. Subsequent developments have expanded this technique in various directions, including ANN-based late-time null tests~\cite{Dialektopoulos:2022late}, the reconstruction of derivatives such as $H'(z)$ together with applications to modified-gravity diagnostics~\cite{Mukherjee:2022hp}, the extension of the reconstruction program to Pantheon and other correlated datasets~\cite{Dialektopoulos:2023pantheon,Dialektopoulos:2023scalar}, and applications to interacting dark-sector phenomenology with the CC and Pantheon$+$ data~\cite{Abedin:2025DarkSector}.

However, a distinct, yet partially overlapping, ANN literature has focused on direct parameter inference rather than function reconstruction. ECoPANN introduced an ANN-based framework for cosmological parameter estimation and demonstrated multi-branch processing of heterogeneous cosmological observables~\cite{Wang:2020ecopann}. Likelihood-free and simulation-based approaches~\cite{Wang:2021likelihoodfree,Wang:2023colfi,Lemos:2023RobustSBI} further extended this direction by combining ANNs with density-estimation strategies to approximate posterior distributions without an explicit analytic likelihood. These studies are important in the present context because they show that ANN architectures can be organized not only as flexible regressors, but also as scalable pipelines for combining multiple probes and for propagating nontrivial parameter degeneracies.
Recent work has also clarified several methodological issues that are directly relevant for ANN reconstruction of the expansion history. One important direction is the treatment of correlated uncertainties. ANN-based reconstructions have been generalized to covariance-aware settings, and more recent CC analyses have shown that covariance information can be incorporated already at the training stage rather than added only at the posterior-analysis stage~\cite{Dialektopoulos:2023pantheon,Chen:2025covmock,Zhang:2024RBFNN}. Another direction is the search for richer architectures. For example, recurrent and hybrid architectures have been explored for cosmological inference. At the same time, Efficient Kolmogorov--Arnold Networks (Ef-KAN) have recently been proposed as a flexible alternative for $H(z)$ reconstruction in strongly nonlinear settings~\cite{Cui:2025lstmefkan}. Hybrid neural--kernel methods have also been proposed to reduce manual kernel selection in GP-style analyses~\cite{Luo:2025NKGPR}. Taken together, these developments indicate that the scientific content of an ANN reconstruction depends not only on the dataset, but also on concrete implementation choices such as the activation function, loss function, training objective, mock-data strategy, covariance treatment, and architecture-selection protocol.

In this manuscript, we propose a framework of an ANN for reconstructing the expansion history of the late-time universe. Unlike ReFANN-type studies, which mainly emphasize direct non-parametric reconstruction from observational data and the subsequent inference of cosmological parameters, and unlike ECoPANN- or CoLFI-type studies, which are primarily designed for fast parameter estimation or likelihood-free posterior inference, our framework is formulated as a reproducible reconstruction pipeline in which the implementation choices themselves are part of the methodology. Specifically, we separate mock-based validation and architecture selection from the final fit to the observed CC sample, so that the activation--architecture pair and the remaining training configuration are fixed before the model is applied to the real data. This staged design is intended to reduce a posteriori tuning to the observations, clarify the impact of the mock-data prescription and training objective on the final reconstruction, and provide a transparent baseline that can later be extended to covariance-aware or multi-probe ANN analyses~\cite{Wang:2020reconstruct,Wang:2020ecopann,Wang:2021likelihoodfree,Wang:2023colfi,Chen:2025covmock,Cui:2025lstmefkan}.

The manuscript is organized as follows. In \zcref{sec:lcdm}, we summarize the flat-$\Lambda$CDM reference adopted for mock generation, describe the CC compilation and preprocessing. In \zcref{sec:validation}, we define the Stage-2 mock-selection statistics and the Stage-3 real-data diagnostics. In \zcref{sec:results}, we present the full-factorial activation--architecture search and the final reconstruction of the observed CC sample. In \zcref{sec:discussion}, we discuss the implications and current limitations of the present baseline. Finally, \zcref{sec:conclusion} summarizes the main conclusions. Throughout this manuscript, explicit values of $H(z)$ are quoted in the conventional units of $\mathrm{km\,s^{-1}\,Mpc^{-1}}$. In general, we use natural units with $c=\hbar=1$, unless conventional observational units are explicitly displayed. We adopt metric signature $(-, +, +, +)$.


\section{\label{sec:lcdm}Methodology}

In this section, we begin summarizing the background explanation of the $\Lambda$CDM because the archived Stage~1 mock generator and several comparison curves used later in this manuscript are defined with respect to this model. Since the target of the present baseline is the Hubble function itself, we retain the derivation up to the redshift-space expression for $H(z)$  explicitly before turning to the CC preprocessing and the staged ANN implementation.

\subsection{\label{subsec:flat_lcdm_reference}Background expansion in flat-$\Lambda$CDM}

Let's start with the Einstein--Hilbert action including the matter, radiation, and a cosmological constant. The action is given by
\begin{equation}
S =
\int d^4x \, \sqrt{-g}
\left[
\frac{R-2\Lambda}{16\pi G} + \mathcal{L}_{\mathrm{m}} + \mathcal{L}_{\mathrm{r}}
\right],
\label{eq:einstein_hilbert_action}
\end{equation}
where $S$ is the total action, $g$ is the determinant of the spacetime metric $g_{\mu\nu}$, $R$ is the Ricci scalar, $\Lambda$ is the cosmological constant, $G$ is Newton's constant, and $\mathcal{L}_{\mathrm{m}}$ and $\mathcal{L}_{\mathrm{r}}$ are the matter and radiation Lagrangian densities, respectively. Varying this action with respect to the metric gives the Einstein equations
\begin{equation}
G_{\mu\nu} + \Lambda g_{\mu\nu} = 8\pi G \, T_{\mu\nu},
\label{eq:einstein_equations_lcdm}
\end{equation}
where $G_{\mu\nu}$ is the Einstein tensor and $T_{\mu\nu}$ is the total energy-momentum tensor. Assuming a homogeneous and isotropic Friedmann--Lema\^itre--Robertson--Walker (FLRW) background,
\begin{equation}
ds^2 = -dt^2 + a^2(t)
\left[
\frac{dr^2}{1-Kr^2} + r^2 d\Omega^2
\right],
\label{eq:flrw_metric}
\end{equation}
where $t$ is cosmic time, $r$ is the comoving radial coordinate, $a(t)$ is the scale factor, $K$ is the spatial-curvature constant, and $d\Omega^2$ is the line element on the unit two-sphere, the $00$ component of \zcref{eq:einstein_equations_lcdm} gives the first Friedmann equation,
\begin{equation}
H^2 + \frac{K}{a^2}
=
\frac{8\pi G}{3}\left(\rho_{\mathrm{m}}+\rho_{\mathrm{r}}+\rho_{\Lambda}\right),
\label{eq:friedmann_first}
\end{equation}
where $H \equiv \dot a / a$ and $\rho_{\Lambda} \equiv \Lambda/(8\pi G)$. The matter, radiation, and vacuum components scale as
\begin{equation}
\rho_{\mathrm{m}}(a)=\rho_{\mathrm{m},0}a^{-3},
\quad
\rho_{\mathrm{r}}(a)=\rho_{\mathrm{r},0}a^{-4},
\quad
\rho_{\Lambda}(a)=\rho_{\Lambda,0},
\label{eq:density_scalings_lcdm}
\end{equation}
where the subscript 0 denotes present-day values and we set $a_0=1$. Introducing the critical density
\begin{equation}
\rho_{\mathrm{c},0} = \frac{3H_0^2}{8\pi G},
\label{eq:critical_density}
\end{equation}
and the standard density parameters
\begin{equation}
\Omega_{i,0} \equiv \frac{\rho_{i,0}}{\rho_{\mathrm{c},0}},
\quad
\Omega_{\Lambda,0} \equiv \frac{\Lambda}{3H_0^2},
\quad
\Omega_{\mathrm{k},0} \equiv -\frac{K}{H_0^2},
\label{eq:omega_definitions_lcdm}
\end{equation}
where the index $i$ labels the matter and radiation components, we can rewrite \zcref{eq:friedmann_first} as
\begin{equation}
H^2(a)
=
H_0^2
\left[
\Omega_{\mathrm{m},0}a^{-3}
+\Omega_{\mathrm{r},0}a^{-4}
+\Omega_{\mathrm{k},0}a^{-2}
+\Omega_{\Lambda,0}
\right].
\label{eq:friedmann_a_form}
\end{equation}
Finally, using the redshift definition $1+z=a^{-1}$, we obtain the redshift-space expansion law
\begin{equation}
H^2(z)
=
H_0^2
\left[
\Omega_{\mathrm{m},0}(1+z)^3
+\Omega_{\mathrm{r},0}(1+z)^4
+\Omega_{\mathrm{k},0}(1+z)^2
+\Omega_{\Lambda,0}
\right],
\label{eq:hfid_stage1}
\end{equation}
with the closure relation
\begin{equation}
\Omega_{\Lambda,0}=1-\Omega_{\mathrm{m},0}-\Omega_{\mathrm{r},0}-\Omega_{\mathrm{k},0}.
\label{eq:omega_lambda_stage1}
\end{equation}

Now, one can use the above expression of the Hubble function $H(z)$ for various cosmological applications. Therefore, we aim to explore the reconstruction technique known as ANN in this study. Before proceeding further, we present a brief overview of the dataset used in the following sections.

\subsection{\label{subsec:data_cc_impl}Cosmic chronometers}

In the present baseline implementation, we focus on the cosmic-chronometer (CC) measurements~\cite{Zhang:2014CC,Simon:2005CC,Moresco:2012CC,Moresco:2016CC6,Ratsimbazafy:2017CC,Stern:2010CC,Borghi:2022CC,Jiao:2023LEGACC,Moresco:2015CC} of the Hubble parameter,
\begin{equation}
H(z) = -\frac{1}{1+z}\frac{dz}{dt},
\label{eq:cc_definition_impl}
\end{equation}
where $z$ is the redshift and $t$ is cosmic time. These measurements provide a late-time estimate of the background expansion rate from the differential-age evolution of passively evolving galaxies.
The raw CC compilation is converted into a standardized table containing the redshift
$z_i$, the observed Hubble rate $H_i \equiv H_{\mathrm{obs}}(z_i)$, and the
reported $1\sigma$ uncertainty $\sigma_{H,i}$. Before the analysis, we verified that the CC dataset used in this study contained no duplicate entries or other evident data-entry errors.
The CC dataset used in the baseline reconstruction can therefore be represented as the following redshift-ordered set
\begin{equation}
\mathcal{D}_{\mathrm{CC}}
=
\left\{
\left(z_i,\,H_i,\,\sigma_{H,i}\right)
\right\}_{i=1}^{N_{\mathrm{CC}}},
\label{eq:cc_dataset}
\end{equation}
with $N_{\mathrm{CC}}=33$ and redshift coverage $0.07 \le z \le 1.965$ in the archived baseline.

In this manuscript, we present the actual observational baseline used in the archived execution. The 33 CC measurements employed in the reconstruction pipeline are listed in \zcref{tab:cc_data_part1,tab:cc_data_part2}.

\begin{table*}[bp]
\caption{\label{tab:cc_data_part1}
Observed CC measurements will be used as the empirical template for the two-stage. In the initial stage, the ReFANN-like mock procedure, calibrates the redshift sampling and the uncertainty model from exactly these measurements.}
\begin{ruledtabular}
\begin{tabular}{cccc}
$z$ & $H(z)$ & $\sigma_H$ & Source \\
 & [km\,s$^{-1}$\,Mpc$^{-1}$] & [km\,s$^{-1}$\,Mpc$^{-1}$] & \\
\hline
0.0700 & 69.00 & 19.600 & Zhang et al.~\cite{Zhang:2014CC} \\
0.0900 & 69.00 & 11.999 & Simon et al.~\cite{Simon:2005CC} \\
0.1200 & 68.60 & 26.200 & Zhang et al.~\cite{Zhang:2014CC} \\
0.1700 & 83.00 & 8.000 & Simon et al.~\cite{Simon:2005CC} \\
0.1791 & 74.91 & 3.807 & Moresco et al.~\cite{Moresco:2012CC} \\
0.1993 & 74.96 & 4.900 & Moresco et al.~\cite{Moresco:2012CC} \\
0.2000 & 72.90 & 29.600 & Zhang et al.~\cite{Zhang:2014CC} \\
0.2700 & 77.00 & 13.999 & Simon et al.~\cite{Simon:2005CC} \\
0.2800 & 88.80 & 36.600 & Zhang et al.~\cite{Zhang:2014CC} \\
0.3519 & 82.78 & 13.948 & Moresco et al.~\cite{Moresco:2012CC} \\
0.3802 & 83.00 & 13.540 & Moresco et al.~\cite{Moresco:2016CC6} \\
0.4000 & 95.00 & 16.995 & Simon et al.~\cite{Simon:2005CC} \\
0.4004 & 76.97 & 10.180 & Moresco et al.~\cite{Moresco:2016CC6} \\
0.4247 & 87.08 & 11.240 & Moresco et al.~\cite{Moresco:2016CC6} \\
0.4497 & 92.78 & 12.900 & Moresco et al.~\cite{Moresco:2016CC6} \\
0.4700 & 89.00 & 49.600 & Ratsimbazafy et al.~\cite{Ratsimbazafy:2017CC} \\
\end{tabular}
\end{ruledtabular}
\end{table*}

\begin{table*}[tbp]
\caption{\label{tab:cc_data_part2} Together, \zcref{tab:cc_data_part1,tab:cc_data_part2} contain the 33-point CC compilation that defines the empirical template used by the archived initial stage and the observational input used by final stage.}
\begin{ruledtabular}
\begin{tabular}{cccc}
$z$ & $H(z)$ & $\sigma_H$ & Source \\
 & [km\,s$^{-1}$\,Mpc$^{-1}$] & [km\,s$^{-1}$\,Mpc$^{-1}$] & \\
\hline
0.4783 & 80.91 & 9.044 & Moresco et al.~\cite{Moresco:2016CC6} \\
0.4800 & 97.00 & 62.002 & Stern et al.~\cite{Stern:2010CC} \\
0.5929 & 103.80 & 12.498 & Moresco et al.~\cite{Moresco:2012CC} \\
0.6797 & 91.60 & 7.962 & Moresco et al.~\cite{Moresco:2012CC} \\
0.7500 & 98.80 & 33.600 & Borghi et al.~\cite{Borghi:2022CC} \\
0.7812 & 104.50 & 12.195 & Moresco et al.~\cite{Moresco:2012CC} \\
0.8000 & 113.10 & 25.220 & Jiao et al.~\cite{Jiao:2023LEGACC}\\
0.8754 & 125.10 & 16.701 & Moresco et al.~\cite{Moresco:2012CC} \\
0.8800 & 90.00 & 39.996 & Stern et al.~\cite{Stern:2010CC} \\
0.9000 & 117.00 & 23.002 & Simon et al.~\cite{Simon:2005CC} \\
1.0370 & 153.70 & 19.674 & Moresco et al.~\cite{Moresco:2012CC} \\
1.3000 & 168.00 & 17.002 & Simon et al.~\cite{Simon:2005CC} \\
1.3630 & 160.00 & 33.580 & Moresco~\cite{Moresco:2015CC} \\
1.4300 & 177.00 & 18.001 & Simon et al.~\cite{Simon:2005CC} \\
1.5300 & 140.00 & 14.000 & Simon et al.~\cite{Simon:2005CC} \\
1.7500 & 202.00 & 39.996 & Simon et al.~\cite{Simon:2005CC} \\
1.9650 & 186.50 & 50.430 & Moresco~\cite{Moresco:2015CC} \\
\end{tabular}
\end{ruledtabular}
\end{table*}

\subsection{\label{subsec:stage_pipeline_cc}Staged workflow for the CC baseline}
A schematic overview of the staged ANN reconstruction framework is presented in \zcref{fig:ann-pipeline}. This figure summarizes the three-stage logic used in the archived baseline process: fiducial mock generation, Stage-2 model-design search, and Stage-3 reconstruction of the observed CC compilation with separate external $H_0$ priors.

\begin{figure}[h]
\centering

\resizebox{\linewidth}{!}{%
\begin{tikzpicture}[node distance=0.8cm and 1.2cm]

  \node[proc] (S1) {Generate 5 mock $H(z)$\\[2pt]datasets from the\\[2pt]fiducial cosmology};

  \node[proc,right=1.4cm of S1] (S2a) {Train the ANN models of\\[2pt]4 activation functions $\times$\\[2pt]12 network architectures\\[2pt]on 5 mocks $\times$ 100 seeds\\[2pt]per candidate};

  \node[diam,right=1.3cm of S2a] (S2b) {Select the best ANN\\[2pt]architecture and\\[2pt]training setup from\\[2pt]repeated-run\\[2pt]risk\,/\,loss\,/\\[2pt]reconstruction\\[2pt]performance};

  \node[proc,right=1.3cm of S2b] (S3a) {Train the ANN model\\[2pt]selected in Stage-2 on CC data\\[2pt]with $H_0$ prior(s) added at $z=0$\\[2pt]and run each prior\\[2pt]case separately};

  \node[para,right=1.2cm of S3a] (S3b) {Output reconstructed\\[2pt]$H(z)$ with an operational\\[2pt]uncertainty band,\\[2pt]flat-$\Lambda$CDM reference,\\[2pt]standardized residuals with respect\\[2pt]to observational uncertainties,\\[2pt]and diagnostic plots};

  \begin{scope}[on background layer]
    \node[stage,fit=(S1),
      label={[font=\small\bfseries\sffamily,text=StageA,anchor=south]north:\textbf{Stage 1}}] {};
    \node[stage,fit=(S2a)(S2b),
      label={[font=\small\bfseries\sffamily,text=StageB,anchor=south]north:\textbf{Stage 2}}] {};
    \node[stage,fit=(S3a)(S3b),
      label={[font=\small\bfseries\sffamily,text=StageC,anchor=south]north:\textbf{Stage 3}}] {};
  \end{scope}

  \draw[arr] (S1)  -- (S2a);
  \draw[arr] (S2a) -- (S2b);
  \draw[arr] (S2b) -- (S3a);
  \draw[arr] (S3a) -- (S3b);

\end{tikzpicture}%
}
\caption{End-to-end pipeline for ANN-based non-parametric reconstruction of the Hubble
parameter $H(z)$. \textbf{Stage~1} generates five CC-like mock $H(z)$ datasets from a fiducial
flat-$\Lambda$CDM cosmology. \textbf{Stage~2} evaluates and selects the best ANN model from 48 combinations of four activation functions and 12 network architectures. Each ANN model candidate is
trained for 100 random initializations on each of the five mocks, giving 500 runs per
candidate and 24,000 runs in total. \textbf{Stage~3} trains the ANN model selected in Stage~2 on the
observed Cosmic Chronometer data with each $H_0$ prior and produces the final reconstructed
$H(z)$ curve with an uncertainty band and diagnostic outputs.}
\label{fig:ann-pipeline}
\end{figure}

In the following, we discuss the work process of our framework.

\begin{itemize}
\item
\textbf{\textit{Stage~1: Mock data generation -} }
A suite of CC-like mock datasets is generated from a fiducial flat-$\Lambda$CDM background. These datasets are used only for controlled validation and design selection, not for the final scientific reconstruction.

\item
\textbf{\textit{Stage~2: Full-factorial ANN design search -} }
The first Stage-2 block trains ANN models consisting of combinations of four activation functions and 12 network architectures under the same weighted-L1 loss function and training settings. Each activation--architecture candidate is trained on all five mock datasets with 100 random initializations per mock.

\item
\textbf{\textit{Stage~2: Candidate aggregation and selection - }} The second block evaluates the performance of the trained ANN model candidates using scores based on mock-truth recovery and smoothness diagnostics.  It then selects the hyperparameter set for the best-performing ANN model, which is used in the next stage for real data.

\item
\textbf{\textit{Stage~3: Observed CC data training - } }
The Stage-2 output is then applied to the observed CC compilation augmented by a Gaussian $H_0$ prior at $z=0$. Separate reconstructions are produced for the Planck 2018, TRGB, and SH0ES R21 priors.

\item
\textbf{\textit{Stage~3: Outputs - }} The reconstructed $H(z)$ and its operational uncertainty band are reported as an ensemble summary over independently initialized training members, combining the mean scale-head output and the member-to-member dispersion.
\end{itemize}

This separation technique between mock-based design and prior-augmented real-data reconstruction is one of the central practical features of the present baseline. It makes the model-selection step auditable, keeps the final real-data application tied to a configuration fixed in advance, and prepares the framework for later extensions to covariance-aware and multi-probe analyses. However, in this baseline, it is important to note that determining the optimal ANN model in Stage-2 can significantly impact the Hubble function reconstruction results using CC data in Stage-3. This topic is discussed in \zcref{subsec:disc_activation_loss}. In the following section, we proceed with the synthetic data generation.

\subsection{\label{subsec:mock_generation_cc}Mock data generation for architecture selection}

For the baseline mock study, we consider the archived fiducial values as
\begin{equation}
H_0 = 70.0,\quad
\Omega_{\mathrm{m},0}=0.3,\quad
\Omega_{\mathrm{k},0}=0,\quad
\Omega_{\mathrm{r},0}=5\times10^{-5},
\label{eq:fiducial_values_stage1}
\end{equation}
where $H_0$ is given in $\mathrm{km\,s^{-1}\,Mpc^{-1}}$. Over the redshift range emphasized by the present CC baseline, the spatial-curvature and radiation contributions are subdominant, and the commonly used flat late-time expression reduces to
\begin{equation}
H(z)
=
H_0
\sqrt{\Omega_{\mathrm{m},0}(1+z)^3+1-\Omega_{\mathrm{m},0}}.
\label{eq:Hz_flat_lcdm}
\end{equation}
In the archived repository, the more general form of \zcref{eq:hfid_stage1} is nevertheless retained in the Stage~1 mock generator so that the same analysis can be reused later for broader mock studies. The reference model is therefore used here as a controlled generator and visual benchmark; so that the selected architecture can be used in the Stage-3 ANN reconstruction itself, which is non-parametric within the class of functions representable by the selected network.

Stage~1 constructs CC-like mock datasets using the fiducial background defined in \zcref{eq:hfid_stage1,eq:fiducial_values_stage1}. For the analysis presented in this work, the generator is operated in a ReFANN-inspired configuration, assuming a fixed flat-$\Lambda$CDM cosmology as the underlying truth. The parameter values are those already specified in \zcref{eq:fiducial_values_stage1}; together with the flatness condition, they imply $\Omega_{\Lambda,0}=0.69995$.

The archived Stage~2 suite is built from five independent mock datasets. By construction, each dataset used by Stage~2 contains 34 points in total: one anchor point at $z=0$ with $H(0)=H_0=70.0\,{\rm km\,s^{-1}\,Mpc^{-1}}$, together with 33 CC-like measurements over the late-time redshift range. Rather than forcing the mock redshifts to coincide with the observed CC redshifts, the generator matches the empirical redshift distribution of the CC sample.

Let $\bar z$ and $s_z^2$ denote the sample mean and variance of the observed redshifts. The mock redshifts are then drawn from a Gamma distribution whose parameters are fixed by moment matching,
\begin{equation}
\alpha = \frac{\bar z^2}{s_z^2},
\quad
\lambda = \frac{\bar z}{s_z^2},
\label{eq:gamma_moments_cc}
\end{equation}
followed by clipping to the target range and sorting in ascending order. Here $\alpha$ is the shape parameter of the Gamma distribution and $\lambda$ is its rate parameter.

The mock uncertainty model is calibrated from the observed CC error bars. First, the reported CC uncertainties are fitted by a linear trend,
\begin{equation}
\bar{\sigma}_H(z)=a_\sigma z+b_\sigma,
\label{eq:sigma_mean_fit}
\end{equation}
where $\bar{\sigma}_H(z)$ is the fitted mean uncertainty at redshift $z$ and $a_\sigma$ and $b_\sigma$ are fitted coefficients. The absolute residual around this trend is then fitted by another linear model,
\begin{equation}
\epsilon_\sigma(z)=a_\epsilon z+b_\epsilon.
\label{eq:sigma_scatter_fit}
\end{equation}
The mock error bar at each redshift is generated as
\begin{equation}
\sigma_H^{\mathrm{mock}}(z)
\sim
\mathcal{N}\!\left(\bar{\sigma}_H(z),\,\epsilon_\sigma^2(z)\right),
\label{eq:sigma_mock_draw}
\end{equation}
and clipped to an empirically motivated interval set by the observed CC uncertainty range.

Finally, the mock Hubble values are drawn as
\begin{equation}
H^{\mathrm{mock}}(z)
=
H_{\mathrm{fid}}(z)
+\delta H(z),
\quad
\delta H(z)\sim \mathcal{N}\!\left(0,\,[\sigma_H^{\mathrm{mock}}(z)]^2\right).
\label{eq:h_mock_draw}
\end{equation}
This design yields a controlled validation set in which the underlying smooth late-time background is known exactly, while the redshift sampling and uncertainty scale remain close to those of the observed CC compilation.
The five generated Stage-1 datasets used in every Stage-2 candidate evaluation are displayed in \zcref{append:mock_datasets}.

\subsection{\label{subsec:network_architecture_cc}Neural-network architecture}

The baseline CC reconstructor is a feed-forward multilayer perceptron that maps
the redshift directly to the Hubble parameter,
\begin{equation}
z \longmapsto H(z).
\label{eq:direct_map_hz}
\end{equation}
The archived Stage-2 and Stage-3 runs do not factorize the prediction as $H(z)=H_0E(z)$; instead, the network directly learns a positive function for $H(z)$ itself.

Before training, the input redshift is normalized as
\begin{equation}
x_i
=
\frac{z_i-z_{\mathrm{loc}}}{z_{\mathrm{scale}}},
\quad
z_{\mathrm{loc}}
=
\frac{z_{\max}+z_{\min}}{2},
\quad
z_{\mathrm{scale}}
=
\max\!\left[
\frac{z_{\max}-z_{\min}}{2},\,10^{-6}
\right],
\label{eq:z_normalization}
\end{equation}
where $z_i$ is the redshift of the $i$th data point, $x_i$ is the corresponding normalized input, and $z_{\min}$ and $z_{\max}$ are the minimum and maximum redshifts in the training set. The target and its observational uncertainty are normalized by a characteristic Hubble scale $h_{\mathrm{scale}}$,
\begin{equation}
y_i = \frac{H_i}{h_{\mathrm{scale}}},
\quad
s_i = \frac{\sigma_{H,i}}{h_{\mathrm{scale}}},
\label{eq:h_normalization}
\end{equation}
where $y_i$ is the normalized target, $s_i$ is the normalized observational uncertainty, and $h_{\mathrm{scale}}$ is taken to be the median of the observed Hubble values in the training sample.

For a network with $L$ hidden layers, the hidden representation is defined by
\begin{align}
\bm{a}^{(0)} &= x,
\\
\bm{u}^{(\ell)} &= \mathbf{W}^{(\ell)}\bm{a}^{(\ell-1)}+\bm{b}^{(\ell)},
\\
\bm{a}^{(\ell)} &= \phi\!\left(\bm{u}^{(\ell)}\right),
\quad \ell=1,\dots,L,
\label{eq:hidden_layers_cc}
\end{align}
where $x$ is the normalized scalar input redshift, $\bm{a}^{(\ell)}$ is the activation vector at hidden layer $\ell$, $\bm{u}^{(\ell)}$ is the corresponding pre-activation vector, $\mathbf{W}^{(\ell)}$ and $\bm{b}^{(\ell)}$ are the trainable weight matrix and bias vector, and $\phi$ denotes the activation function. The archived search compares one- and two-hidden-layer models, with candidate widths ranging from 8 to 256 neurons.

The output layer predicts a positive Hubble rate through a softplus transform,
\begin{equation}
\hat y_i
=
\mathrm{softplus}\!\left(o_{\theta}(x_i)\right)+\varepsilon_H,
\quad
\hat H_i
=
h_{\mathrm{scale}}\,\hat y_i,
\label{eq:positive_output_h}
\end{equation}
where $o_{\theta}(x_i)$ is the scalar pre-activation output of the network with trainable parameter set $\theta$ evaluated at the normalized input $x_i$, $\hat y_i$ is the normalized positive prediction, $\hat H_i$ is the corresponding dimensional prediction, and $\varepsilon_H>0$ is a small numerical floor.
In the archived Stage-2 and Stage-3 runs, the network also uses a second positive output head for a scale prediction,
\begin{equation}
\hat s_{\mathrm{scale},i}
=
\mathrm{softplus}\!\left(o_{\theta}^{(\sigma)}(x_i)\right)+\varepsilon_\sigma,
\quad
\hat{\sigma}_{\mathrm{scale},i}
=
h_{\mathrm{scale}}\,\hat s_{\mathrm{scale},i},
\label{eq:positive_output_sigma}
\end{equation}
where $\hat{\sigma}_{\mathrm{scale},i}$ denotes the positive scale-head output produced jointly with the mean reconstruction. The plotted Stage-2 and Stage-3 bands distinguish this output from the member-to-member ensemble spread introduced at the final reconstruction stage.

\subsection{\label{subsec:losses_cc}Loss functions implemented in the framework}

The framework can accommodate several robust losses for one-dimensional Hubble-rate reconstruction, but the archived Stage-1--Stage-3 results reported in this manuscript use the \emph{weighted-L1} loss function throughout the design study and the final CC reconstruction. For normalized targets, the data term takes the form
\begin{equation}
\mathcal{L}_{\mathrm{wL1}}
=
\frac{1}{N}
\sum_{i=1}^{N}
\frac{\left|\hat y_i-y_i\right|}{s_i+\varepsilon_s},
\label{eq:weighted_l1_cc}
\end{equation}
where $\hat y_i$ is the normalized ANN prediction, $y_i$ is the normalized target, $s_i$ is the normalized observational uncertainty, and $\varepsilon_s$ is a small numerical floor.

The practical point for our work is that the Stage-2 design comparison is carried out at fixed loss function while activation function and network architecture are varied jointly. The complete search over all activation--architecture combinations therefore retains their interactions instead of estimating the two effects in a sequential screening procedure. The scale head remains active in the archived runs, while the final Stage-3 band additionally incorporates the ensemble dispersion across independently initialized members.

\subsection{\label{subsec:training_selection_cc}Training, exhaustive design search, and final ensemble}

For all archived Stage-2 and Stage-3 runs discussed in this manuscript, the trainable parameters are optimized with the Adam optimizer. The learning rate is fixed to $10^{-3}$, the maximum number of epochs is $800$, the gradient norm is clipped at $5.0$, dropout is set to zero, and the minimum improvement threshold for early stopping is $10^{-4}$. The stage-design runs use no validation split, so the comparison is driven by repeated mock datasets and repeated random initializations rather than by a single train/validation partition.

Stage~2 performs one exhaustive design search rather than a sequential activation screening and architecture scan. The activation set is the Exponential Linear Unit (ELU)~\cite{2015arXiv151107289C}, the hyperbolic tangent ($\tanh$)~\cite{LeCun1998}, the Sigmoid-weighted Linear Unit (SiLU)~\cite{ELFWING20183}, and the Tanh Exponential (TanhExp)~\cite{liu2021tanhexp}. Each activation is crossed with six one-hidden-layer structures and six equal-width two-hidden-layer structures,
\begin{equation}
\begin{aligned}
\mathcal{A}
&=
\left\{\mathrm{ELU},\mathrm{SiLU},\tanh,\mathrm{TanhExp}\right\},
\\
\mathcal{C}
&=
\left\{[n],[n,n]\;\middle|\;n\in\{8,16,32,64,128,256\}\right\}.
\end{aligned}
\label{eq:stage2_full_factorial_grid}
\end{equation}
The $4\times12$ activation--structure combinations defined by $\mathcal{A}$ and $\mathcal{C}$ therefore yield 48 candidates. Every candidate is evaluated on five independent mock datasets and 100 random initializations per mock, giving 500 runs per candidate and 24,000 Stage-2 training runs in total. The weighted-L1 loss function, active scale head, optimizer, and all other training settings are held fixed across the grid.

Because the mock truth is known at Stage~2, each activation--architecture candidate is selected by a
truth-aware score rather than by the real-data loss alone.
The primary term is
\begin{equation}
S_{\mathrm{select}}
=
\frac{1}{N}
\sum_{i=1}^{N}
\left[
\frac{\hat H(z_i)-H_{\mathrm{true}}(z_i)}{\sigma_{H,i}}
\right]^2,
\label{eq:s_select}
\end{equation}
where $N$ is the number of evaluation redshift points, $\hat H(z_i)$ is the reconstructed Hubble rate, $H_{\mathrm{true}}(z_i)$ is the fiducial mock truth, and $\sigma_{H,i}$ is the mock observational uncertainty. This quantity measures the truth-recovery error in units of the mock observational uncertainty.
To suppress spurious oscillatory solutions, we also define a smoothness penalty
using a finite-difference estimate of the second derivative of the normalized reconstructed curve, where the median is taken over the absolute values of the reconstructed curve on the evaluation grid,
\begin{equation}
R_{\mathrm{smooth}}
=
\frac{1}{N}
\sum_{i=1}^{N}
\left[
\frac{d^2}{dz^2}
\left(
\frac{\hat H(z)}{\mathrm{median}\,|\hat H|}
\right)_{z=z_i}
\right]^2.
\label{eq:r_smooth}
\end{equation}
The per-run candidate ranking statistic is then
\begin{equation}
\mathrm{Score}
=
S_{\mathrm{select}}
+
\lambda_{\mathrm{smooth}}R_{\mathrm{smooth}},
\quad
\lambda_{\mathrm{smooth}}=0.05.
\label{eq:architecture_score}
\end{equation}
Additional diagnostics such as the reduced $\chi^2$ with respect to the mock
observations, the Root Mean Square Error (RMSE) with respect to the truth, the empirical 68\% coverage,
and the normalized residual standard deviation are recorded for interpretation.
The per-run statistic in \zcref{eq:architecture_score} is then aggregated over mock datasets and random initializations according to the Stage-2 rules defined in \zcref{subsec:stage2_validation}.

After the Stage-2 design study is completed, Stage~3 trains the selected architecture on the observed CC compilation augmented by one external $H_0$ prior at $z=0$. The archived Stage-3 execution consists of 100 independently initialized members for each of the three prior choices. For a fixed prior choice $k$, the ensemble mean and standard deviation at each redshift are
\begin{equation}
\bar H^{(k)}(z_i)
=
\frac{1}{M}
\sum_{m=1}^{M}
\hat H^{(k)}_{m}(z_i),
\label{eq:ensemble_mean}
\end{equation}
\begin{equation}
\sigma_{\mathrm{ens}}^{(k)}(z_i)
=
\sqrt{
\frac{1}{M-1}
\sum_{m=1}^{M}
\left[
\hat H^{(k)}_m(z_i)-\bar H^{(k)}(z_i)
\right]^2
},
\label{eq:ensemble_std}
\end{equation}
with $M=100$ in the archived run. The member-level scale-head outputs are averaged as
\begin{equation}
\bar\sigma_{\mathrm{head}}^{(k)}(z_i)
=
\frac{1}{M}
\sum_{m=1}^{M}
\hat\sigma_{\mathrm{scale},m}^{(k)}(z_i).
\label{eq:ensemble_sigma_head_mean}
\end{equation}
For the operational Stage-3 band and residual diagnostics, this mean scale-head output and the ensemble dispersion are combined in quadrature,
\begin{equation}
\sigma_{\mathrm{tot}}^{(k)}(z_i)
=
\sqrt{
\left[\bar\sigma_{\mathrm{head}}^{(k)}(z_i)\right]^2
+
\left[\sigma_{\mathrm{ens}}^{(k)}(z_i)\right]^2
}.
\label{eq:sigma_total_cc}
\end{equation}
Accordingly, the Stage-3 bands shown below are not dropout-based bands, single-model error bars, or posterior credible intervals; they are explicit operational combinations of the mean scale-head output and the dispersion across the archived 100-member runs.

\section{\label{sec:validation}Validation strategy and evaluation metrics}

In this section, we define the quantitative diagnostics used to assess the mock-based selection stage and the final real-data reconstruction. This manuscript is organized around two distinct validation regimes. Stage~2 is a controlled mock-based experiment in which the true background is known and can therefore be used explicitly in model selection. Stage~3, by contrast, is the reconstruction of the observed CC sample, for which only data-level residual diagnostics are available. The quantitative criteria used at the two stages are therefore related but not identical.

\subsection{\label{subsec:stage2_validation}Stage 2 metrics for mock-based selection}

For every activation--architecture candidate in the full grid of \zcref{eq:stage2_full_factorial_grid}, the central ranking statistic is the mock-truth score of \zcref{eq:architecture_score}.

Let $q=(a,c)$ denote one of the 48 candidates, where $a\in\mathcal{A}$ is the activation function and $c\in\mathcal{C}$ is the network structure. Let $m=1,\ldots,M$ index the mock dataset and $s=1,\ldots,N_s$ the random initialization, with $M=5$ and $N_s=100$. The per-run score is denoted by
\begin{equation}
\mathrm{Score}_{q,m,s}
=
S_{\mathrm{select},q,m,s}
+
\lambda_{\mathrm{smooth}}R_{\mathrm{smooth},q,m,s}.
\label{eq:stage2_activation_per_run_score}
\end{equation}
To reduce sensitivity to occasional failed initializations while retaining the mock-dataset dependence of the Stage-2 suite, the implementation first summarizes the seed direction by
\begin{equation}
\widetilde{\mathrm{Score}}_{q,m}
=
\operatorname{median}_{s}\left(\mathrm{Score}_{q,m,s}\right),
\label{eq:stage2_activation_seed_median}
\end{equation}
and then averages the five mock-level seed medians,
\begin{equation}
\mathcal{J}_{q}
=
\frac{1}{M}
\sum_{m=1}^{M}
\widetilde{\mathrm{Score}}_{q,m}.
\label{eq:stage2_activation_statistic}
\end{equation}
The primary export rule minimizes this hierarchical statistic over the complete activation--architecture grid,
\begin{equation}
q_\star
=
\operatorname*{arg\,min}_{q\in\mathcal{A}\times\mathcal{C}} \mathcal{J}_{q}.
\label{eq:stage2_activation_argmin}
\end{equation}
For sensitivity to the aggregation order, we additionally record the pooled median over all mock--seed runs,
\begin{equation}
\widetilde{\mathrm{Score}}_{q}^{\mathrm{pool}}
=
\operatorname{median}_{m,s}\left(\mathrm{Score}_{q,m,s}\right),
\quad
q_{\star}^{\mathrm{pool}}
=
\operatorname*{arg\,min}_{q\in\mathcal{A}\times\mathcal{C}}
\widetilde{\mathrm{Score}}_{q}^{\mathrm{pool}}.
\label{eq:stage2_architecture_export_rule}
\end{equation}
The candidate that minimizes $\mathcal{J}_q$ is exported to Stage~3, whereas the minimizer of the pooled median is retained only as an aggregation-sensitivity check. For each candidate, we also record the mean score, the mean truth-recovery term $S_{\mathrm{select}}$, the mean smoothness penalty $R_{\mathrm{smooth}}$, the reduced $\chi^2$ with respect to the known truth, the RMSE with respect to the true background, the empirical 68\% coverage, and the training-loss diagnostics. The resulting candidate rankings and selected configuration are reported in \zcref{subsec:results_stage2}.

\subsection{\label{subsec:stage3_metrics}Stage 3 diagnostics for the observed CC sample}

For the Stage-3 real-data runs, the CC compilation is augmented by a single Gaussian prior on the present-day Hubble rate,
\begin{equation}
\mathcal{D}_{\mathrm{CC}+H_0}^{(k)}
=
\mathcal{D}_{\mathrm{CC}}
\cup
\left\{
\left(z=0,\;H_0^{(k)},\;\sigma_{H_0}^{(k)}\right)
\right\},
\label{eq:cc_h0_augmented_dataset}
\end{equation}
where the superscript $k$ labels the adopted external $H_0$ prior. In the archived Stage-3 execution, three separate prior choices are considered: the Planck 2018 prior, the TRGB prior, and the SH0ES R21 prior. Each Stage-3 run therefore fits $N_{\mathrm{CC}}+1=34$ data points.

At Stage~3 the truth is no longer known, so the diagnostics are defined relative to the prior-augmented observational sample $\mathcal{D}_{\mathrm{CC}+H_0}^{(k)}$ for each prior choice $k$. For the final ensemble mean prediction $\bar H^{(k)}(z_i)$, we compute the RMSE, Mean Absolute Error (MAE), signed bias, residual standard deviation, and the mean relative signed and absolute errors at the sampled redshifts.

The residual standardized by the total operational scale is defined as
\begin{equation}
r_i^{\mathrm{norm},(k)}
=
\frac{H_i-\bar H^{(k)}(z_i)}{\sigma_{\mathrm{tot}}^{(k)}(z_i)},
\label{eq:norm_resid_stage3_real}
\end{equation}
where $\sigma_{\mathrm{tot}}^{(k)}$ is the total operational scale of \zcref{eq:sigma_total_cc}. We monitor the mean and standard deviation of $r_i^{\mathrm{norm},(k)}$. For an interval probability $p$, we also define
\begin{equation}
\begin{aligned}
C_p^{(k)}
&=
\frac{1}{N}
\sum_{i=1}^{N}
\mathbf{1}\!\left(\left|r_i^{\mathrm{norm},(k)}\right|\le q_p\right),
\\
q_p
&=
\Phi^{-1}\!\left(\frac{1+p}{2}\right).
\end{aligned}
\label{eq:stage3_in_sample_containment}
\end{equation}
where $\Phi^{-1}$ is the standard-normal quantile function. Because the same 34 observations are used both to train the ensemble and to evaluate \zcref{eq:stage3_in_sample_containment}, $C_{68}$ and $C_{95}$ are in-sample containment fractions. They are not coverage estimates for the latent expansion history or future observations. The residuals standardized by $\sigma_{\mathrm{tot}}$ are distinct from those displayed in the Stage-3 Hubble figures, for which the denominator is the reported observational uncertainty $\sigma_{\mathrm{data}}$. Since the three Stage-3 runs only differ in the external $H_0$ datum at $z=0$, these diagnostics illustrate how the low-redshift anchor is incorporated into the late-time reconstruction fit.

We also calculate the $O\mathrm{m}(z)$ diagnostic~\cite{Sahni:2008Om},
\begin{equation}
O\mathrm{m}^{(k)}(z)
=
\frac{
\left[\bar H^{(k)}(z)/H_{0,\mathrm{ANN}}^{(k)}\right]^2-1
}
{(1+z)^3-1},
\label{eq:om_diagnostic_stage3}
\end{equation}
where $H_{0,\mathrm{ANN}}^{(k)}=\bar H^{(k)}(0)$ is the saved Stage-3 ANN prediction at the unique $z=0$ anchor row. The horizontal comparison line is the matter-density parameter $\Omega_{\mathrm{m},0}^{\Lambda\mathrm{CDM},(k)}$ obtained from the flat-$\Lambda$CDM fit to the same prior-augmented dataset. In a spatially flat $\Lambda$CDM model the diagnostic is constant and equal to $\Omega_{\mathrm{m},0}$. The uncertainty bands in panel (b) of the Stage-3 figures are propagated from the ANN Hubble-rate uncertainty as
\begin{equation}
\sigma_{O\mathrm{m}}^{(k)}(z)
=
\left|
\frac{2\bar H^{(k)}(z)}
{\left[H_{0,\mathrm{ANN}}^{(k)}\right]^2\left[(1+z)^3-1\right]}
\right|
\sigma_H^{(k)}(z),
\label{eq:om_uncertainty_stage3}
\end{equation}
where $\sigma_H^{(k)}(z)$ denotes the total operational scale used for the Stage-3 diagnostic plot. Since the denominator of \zcref{eq:om_diagnostic_stage3} vanishes at $z=0$, the singular point is masked in the visualization. The lower panel of each $O\mathrm{m}(z)$ plot is therefore used only as a comparison-focused zoomed view of the same diagnostic.

\section{\label{sec:results}Reconstruction results of $H(z)$ from the CC dataset}

In this section, we present the results obtained from the archived CC baseline implementation after applying the mock-level aggregation rules described in \zcref{subsec:stage2_validation}. Unless otherwise stated, the Stage-2 numbers quoted below are aggregates over the full Stage-2 suite, namely five independent CC-like mock datasets and 100 random initializations for each network setting, for a total of 500 runs per configuration.

\subsection{\label{subsec:results_stage2}Stage-2: full-factorial activation--architecture search}

In Stage-2, we evaluate all 48 candidates of the activation-architecture ANN model under the common weighted-L1 loss function and specified training settings. Each candidate contributes 500 runs, comprising five mock datasets and 100 random initializations per mock. As a result of this evaluation, the primary hierarchical statistic of \zcref{eq:stage2_activation_statistic} is minimized by SiLU-$[8]$, with $\mathcal{J}=0.0473$. The secondary pooled-median statistic is minimized by SiLU-$[16]$, with $\widetilde{\mathrm{Score}}^{\mathrm{pool}}=0.0473$. However, it should be noted that the selection itself utilizes the distinct full-precision values (for a comprehensive overview of the scores for all 48 candidates, refer to the \zcref{tab:stage2_full_factorial} in the \zcref{append:score_of_stage2-design}). The executed Stage-3 configuration follows the primary rule and therefore uses SiLU-$[8]$.

\begin{figure*}[tbp]
  \centering
  \subfloat[ELU\label{fig:stage2-elu-depth1}]{%
    \includegraphics[width=0.492\linewidth]{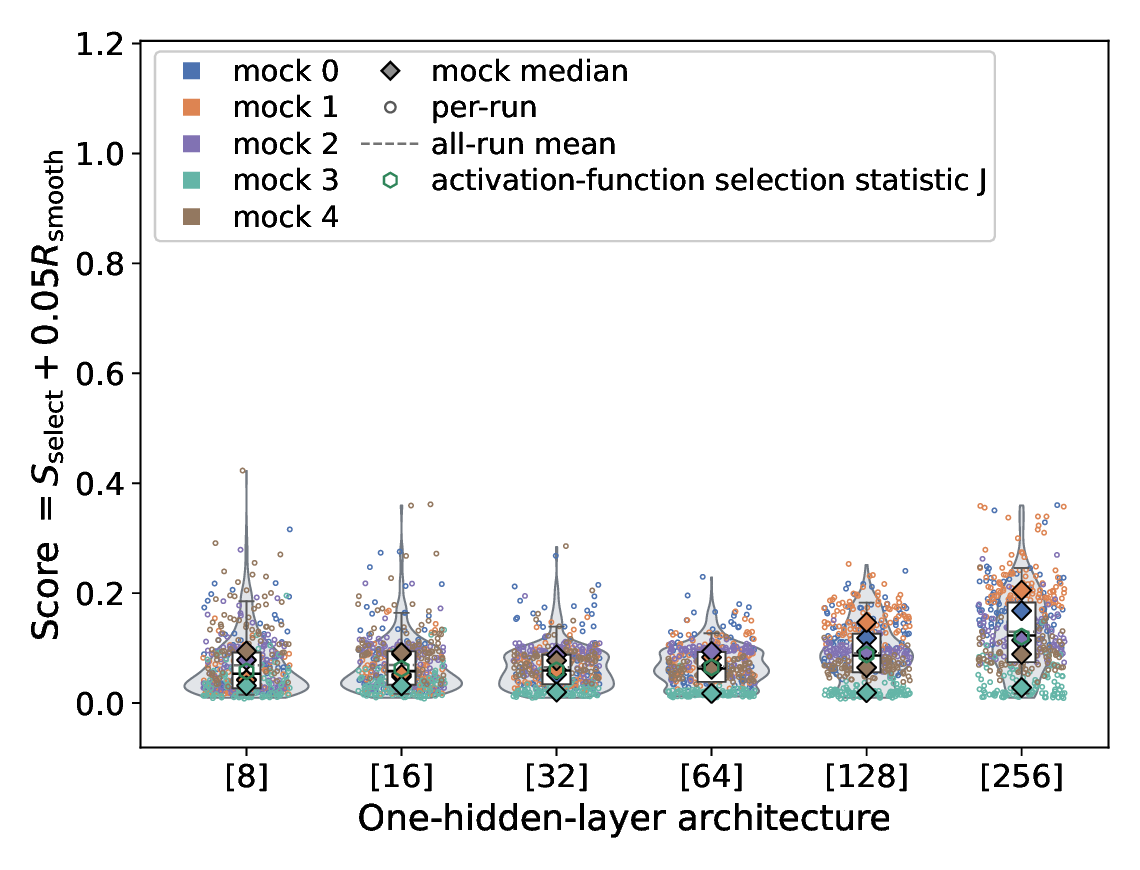}
  }\hfill
  \subfloat[SiLU\label{fig:stage2-silu-depth1}]{%
    \includegraphics[width=0.492\linewidth]{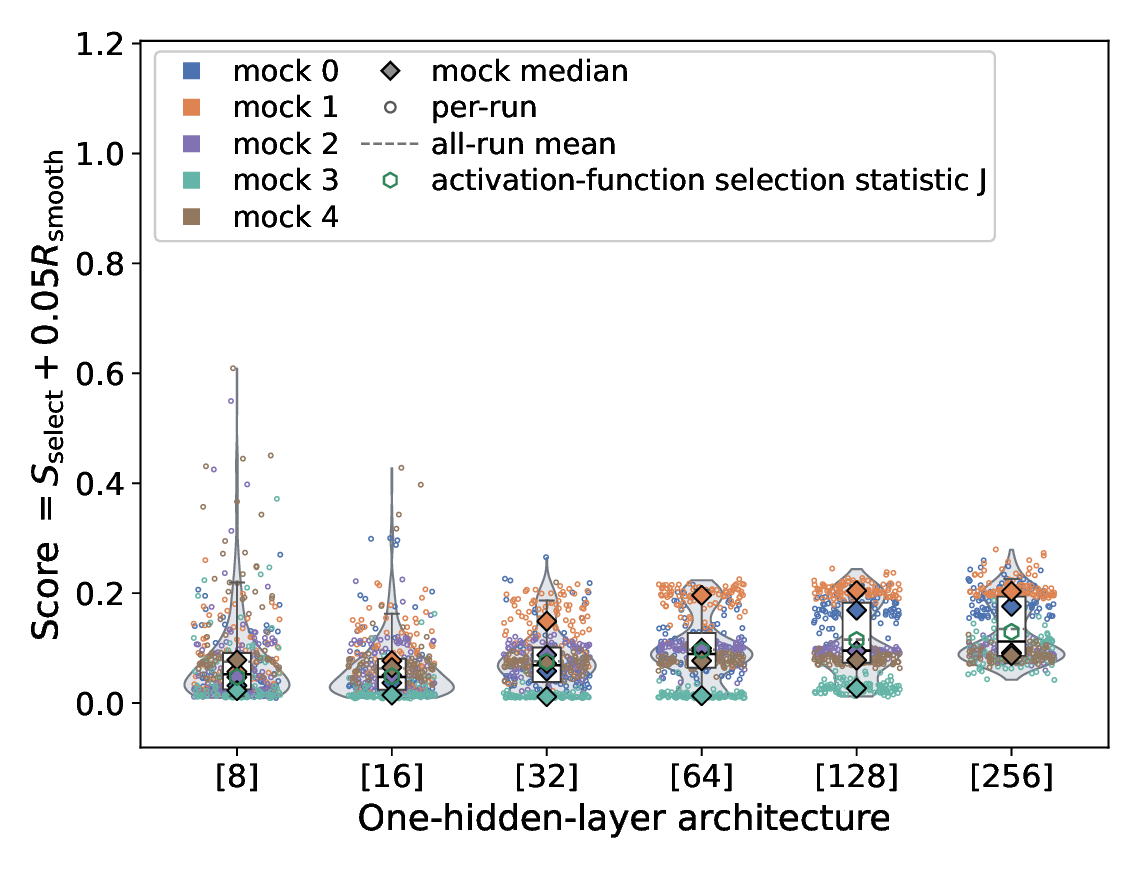}
  }\\
  \subfloat[$\tanh$\label{fig:stage2-tanh-depth1}]{%
    \includegraphics[width=0.492\linewidth]{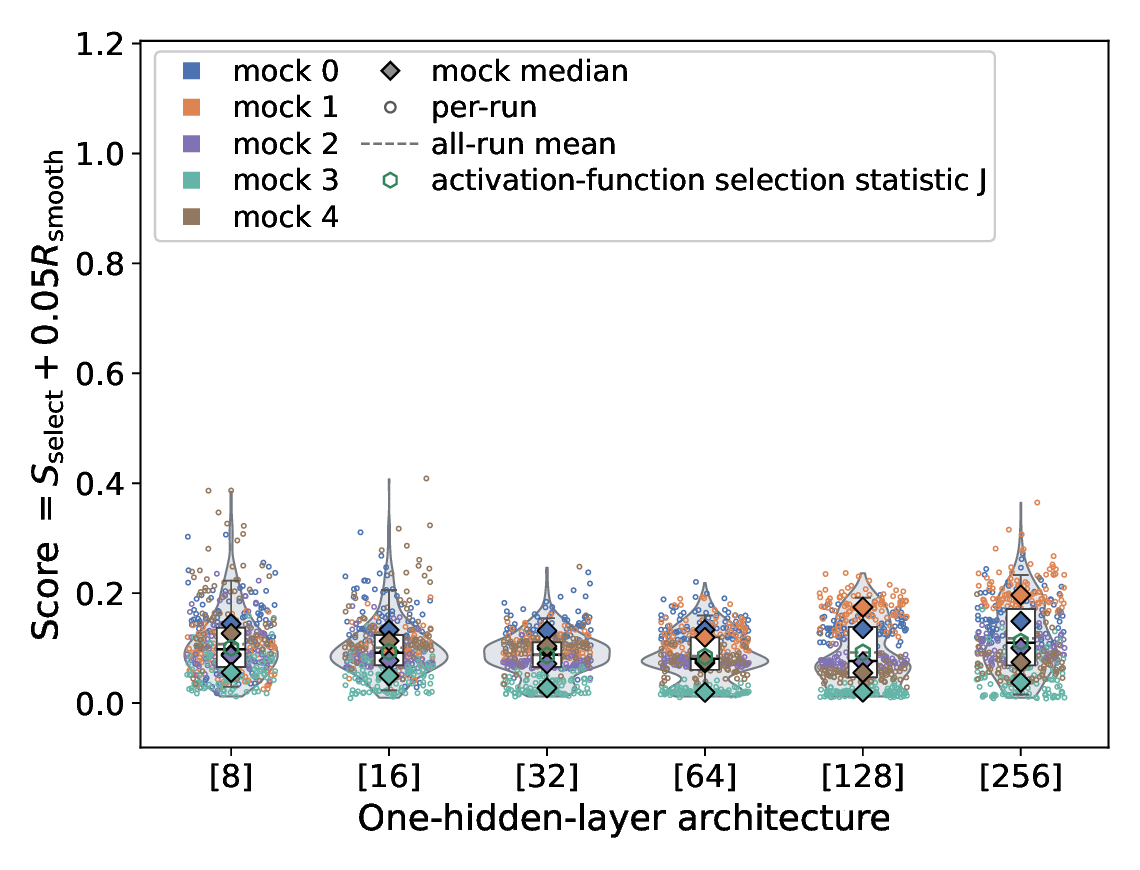}
  }\hfill
  \subfloat[TanhExp\label{fig:stage2-tanhexp-depth1}]{%
    \includegraphics[width=0.492\linewidth]{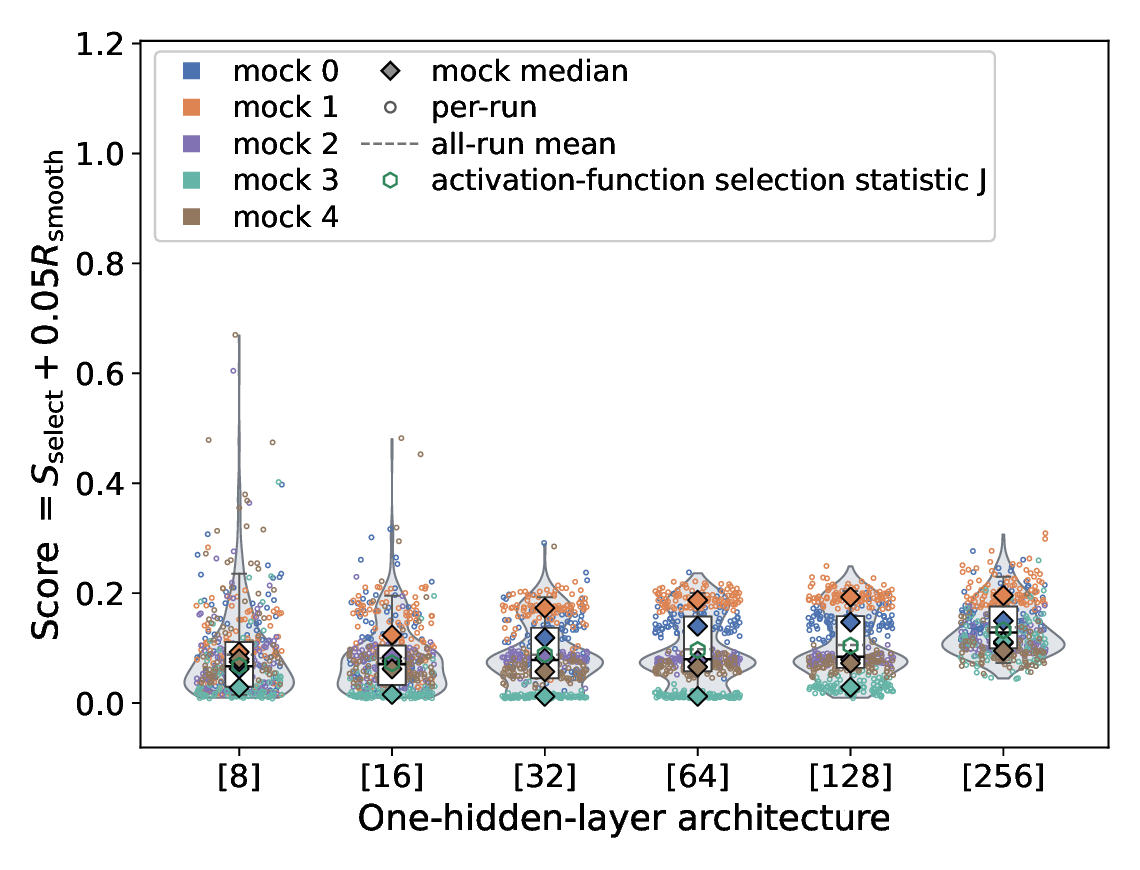}
  }
  \caption{\label{fig:stage2_score_depth1}
  Stage-2 Score distributions for the six one-hidden-layer structures $[8]$, $[16]$, $[32]$, $[64]$, $[128]$, and $[256]$, shown separately for (a) ELU, (b) SiLU, (c) $\tanh$, and (d) TanhExp. Each architecture contains 500 run-level Scores (five mocks $\times$ 100 seeds) from the full-factorial search; lower values are preferred. The violin and internal box summarize the pooled distribution, with whiskers at the 5th and 95th percentiles. Colored open circles are individual runs, diamonds are the seed medians for mocks 0--4, the gray dashed segment is the all-run mean, and the green outlined marker is the hierarchical statistic $\mathcal{J}$, i.e. the mean of the five mock-specific seed medians. All cases use weighted L1 loss, zero dropout, learning rate $10^{-3}$, maximum epoch number 800, and no validation split. The minimum $\mathcal{J}$ within each activation panel is ELU-$[8]$: $0.0577$; SiLU-$[8]$: $0.0473$; $\tanh$-$[64]$: $0.0844$; and TanhExp-$[8]$: $0.0677$. Thus SiLU-$[8]$ is also the global minimum under the primary selection rule.}
\end{figure*}

\begin{figure*}[tbp]
  \centering
  \subfloat[ELU\label{fig:stage2-elu-depth2}]{%
    \includegraphics[width=0.492\linewidth]{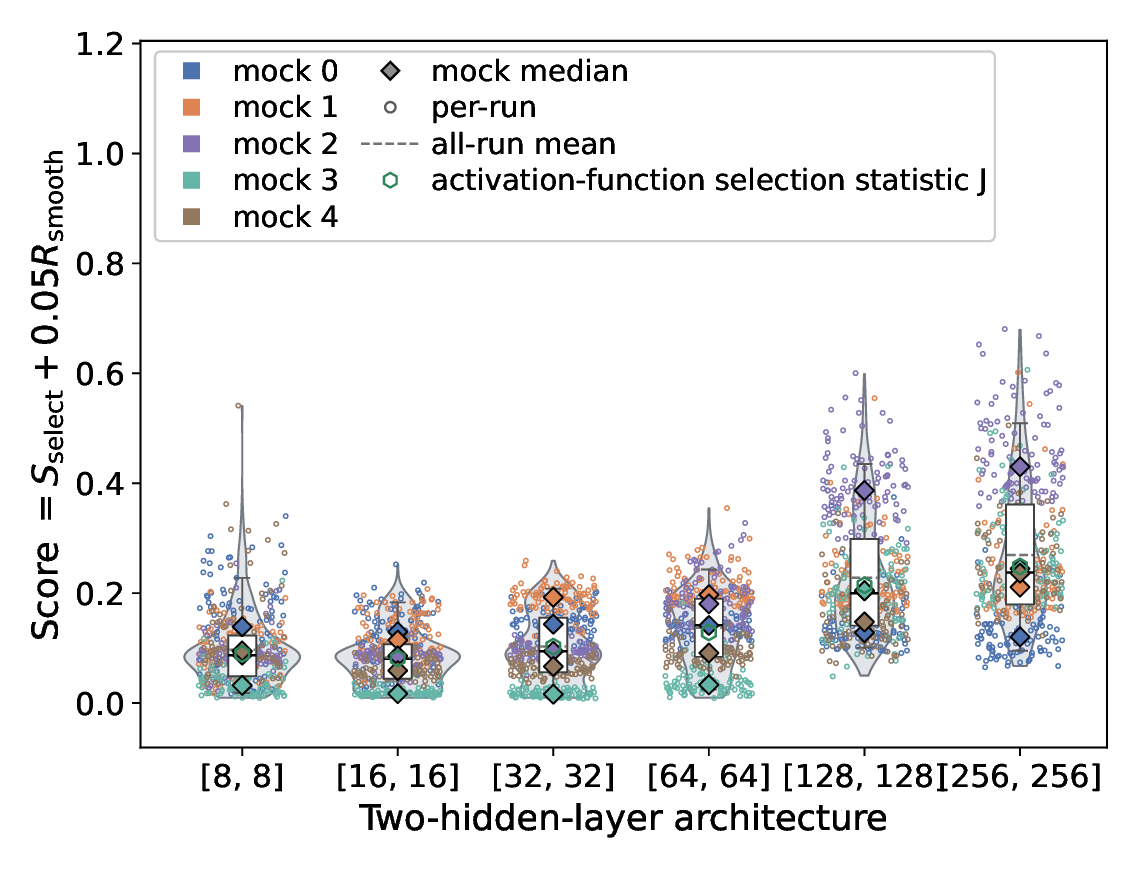}
  }\hfill
  \subfloat[SiLU\label{fig:stage2-silu-depth2}]{%
    \includegraphics[width=0.492\linewidth]{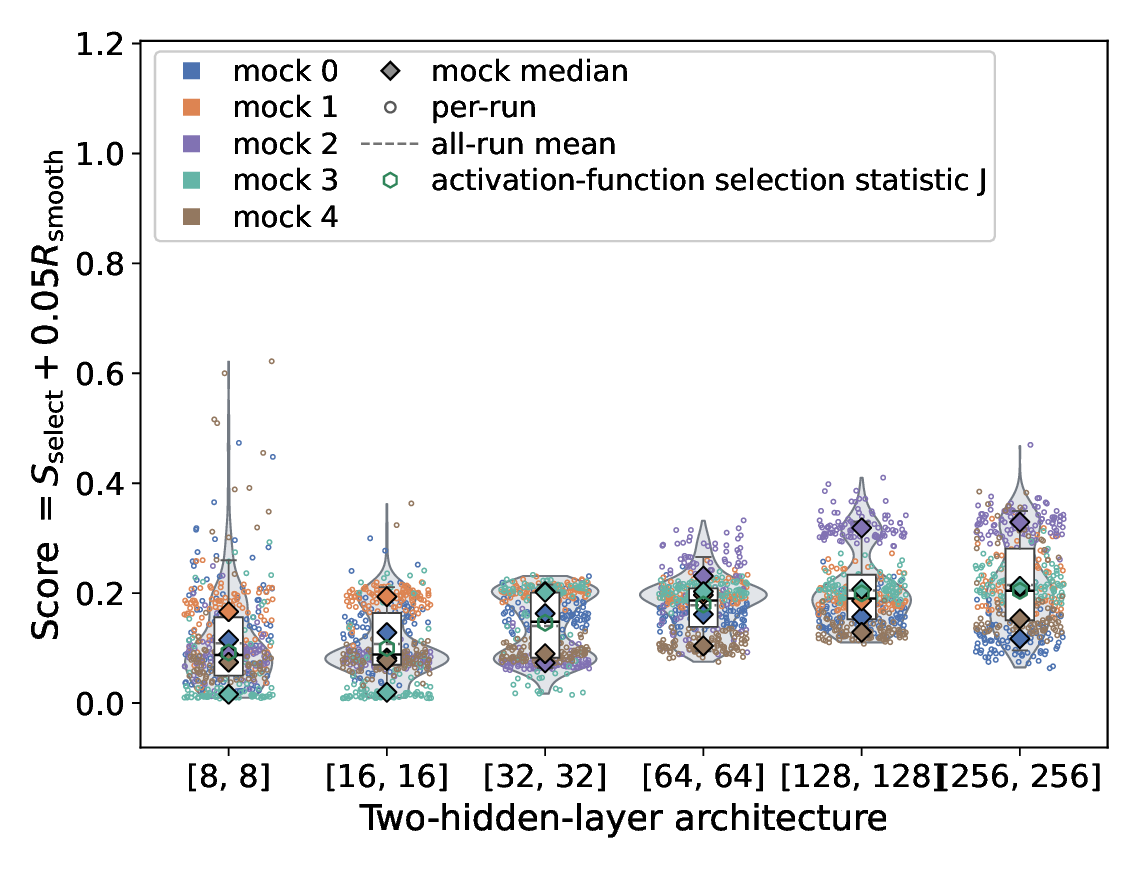}
  }\\
  \subfloat[$\tanh$\label{fig:stage2-tanh-depth2}]{%
    \includegraphics[width=0.492\linewidth]{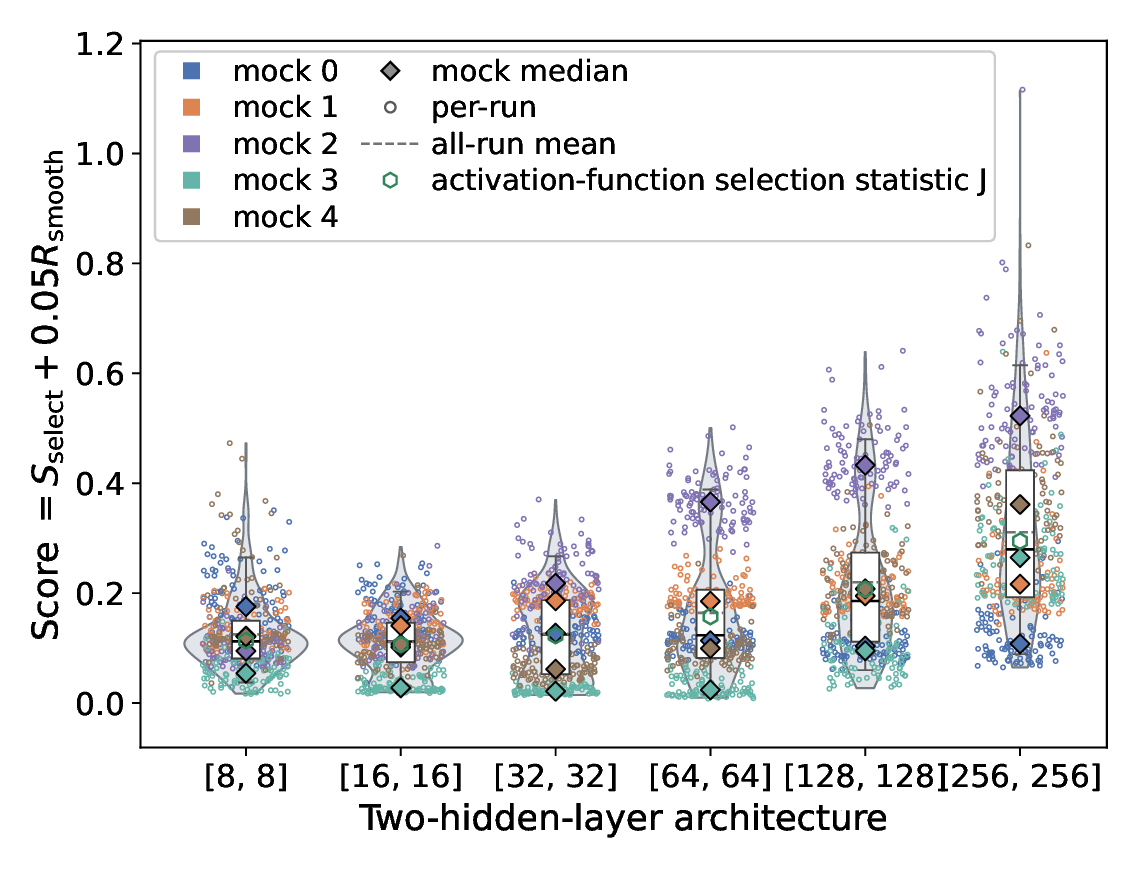}
  }\hfill
  \subfloat[TanhExp\label{fig:stage2-tanhexp-depth2}]{%
    \includegraphics[width=0.492\linewidth]{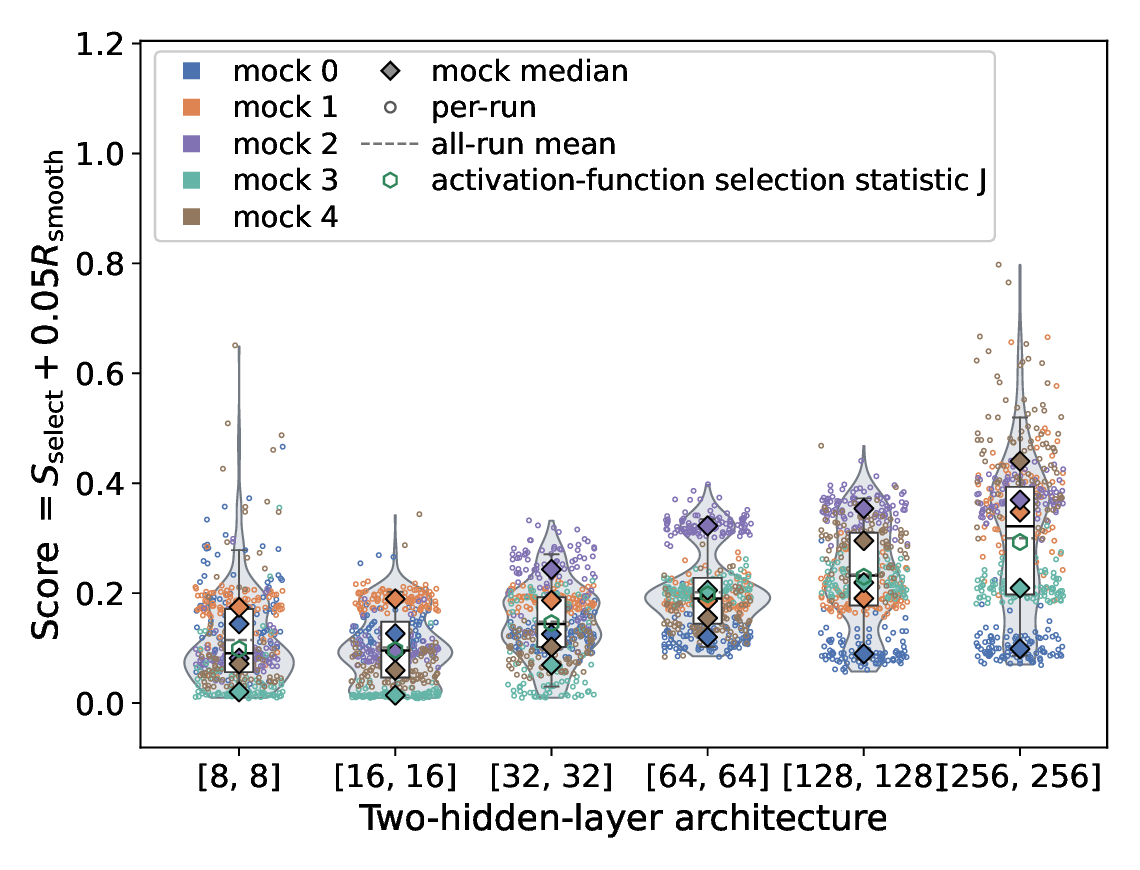}
  }
  \caption{\label{fig:stage2_score_depth2}
  Stage-2 Score distributions for the six equal-width two-hidden-layer structures $[8,8]$, $[16,16]$, $[32,32]$, $[64,64]$, $[128,128]$, and $[256,256]$, shown separately for (a) ELU, (b) SiLU, (c) $\tanh$, and (d) TanhExp. Each architecture contains 500 run-level Scores, and the violin, box, 5th--95th percentile whiskers, per-run points, mock-specific seed-median diamonds, all-run mean, and hierarchical $\mathcal{J}$ marker follow the conventions of \zcref{fig:stage2_score_depth1}. All cases use weighted L1 loss, zero dropout, learning rate $10^{-3}$, maximum epoch number 800, and no validation split. The minimum $\mathcal{J}$ within each activation panel is ELU-$[16,16]$: $0.0812$; SiLU-$[8,8]$: $0.092$; $\tanh$-$[16,16]$: $0.106$; and TanhExp-$[16,16]$: $0.0964$. All four depth-two minima exceed the global full-grid minimum $\mathcal{J}=0.0473$ of SiLU-$[8]$.}
\end{figure*}

Here, we present violin plots of the score distributions for each run of the respective ANN models, separated into one-layer and two-layer cases, as shown in \zcref{fig:stage2_score_depth1,fig:stage2_score_depth2}. The one- and two-hidden-layer families show different width dependence. For the one-hidden-layer ANN models in \zcref{fig:stage2_score_depth1}, changes in the width of the setting, or the number of nodes per model, are comparatively gradual, and the four activation functions do not show a common sharp transition over most of the range. At $[256]$, shifts in the mock-specific point concentrations become more visible, particularly for SiLU and TanhExp. Therefore, the apparent stability at smaller widths should not be mentioned outside of the width range. For the two-hidden-layer models in \zcref{fig:stage2_score_depth2}, the Score distributions generally move upward and develop broader upper tails as the width increases. The onset differs by activation: the change is already visible near $[32,32]$ for SiLU and TanhExp, becomes clear by approximately $[64,64]$ for ELU, and appears mainly as broadening near $[64,64]$ followed by a stronger upward shift for the wider $\tanh$ models. Thus, increasing both depth and width tends to reduce operational stability, where ``less stable'' refers to larger and more dispersed Scores rather than to failed numerical convergence.

The mock-colored points and mock-specific median diamonds also clarify what can and cannot be inferred from a violin. A tight vertical concentration of points of the same color indicates weak seed dependence for that particular mock, whereas a large vertical spread within one color indicates stronger seed-to-seed variation. The horizontal width of the violin at a given Score represents estimated density and is not itself an error bar. Closely aligned mock-median diamonds indicate limited mock-to-mock sensitivity of the Score, while separated diamonds indicate that the ranking statistic depends on the mock dataset. Such separation can reflect differences in redshift sampling, noise, or intrinsic reconstruction difficulty; by itself it does not establish overfitting. Likewise, the non-negative truth-recovery Score does not retain the sign needed to diagnose prediction bias. A formal bias, variance, or overfitting assessment necessitates an independent evaluation set or additional mocks.

\begin{figure*}[tbp]
  \centering
  \subfloat[ELU-$\lbrack128\rbrack$: Score $=0.00773$\label{fig:stage2-best-elu}]{%
    \includegraphics[width=0.492\linewidth]{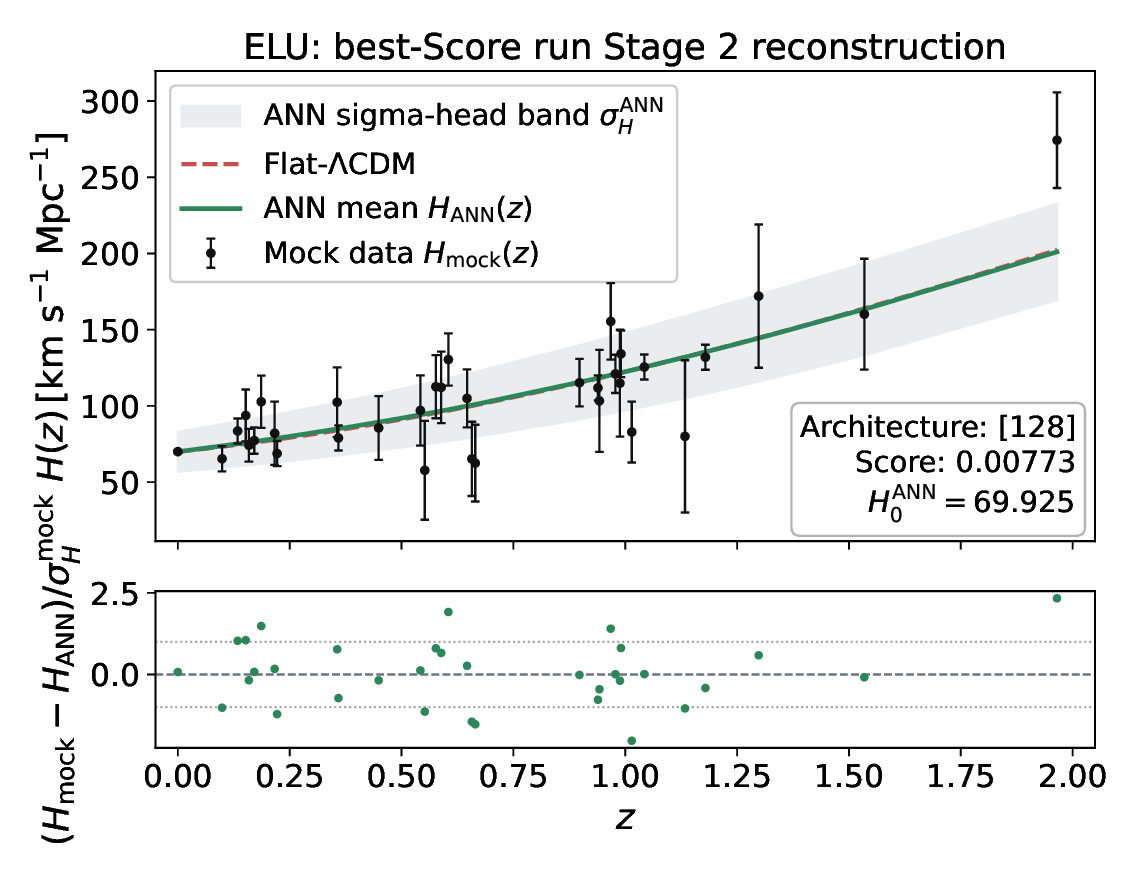}
  }\hfill
  \subfloat[SiLU-$\lbrack8,8\rbrack$: Score $=0.00814$\label{fig:stage2-best-silu}]{%
    \includegraphics[width=0.492\linewidth]{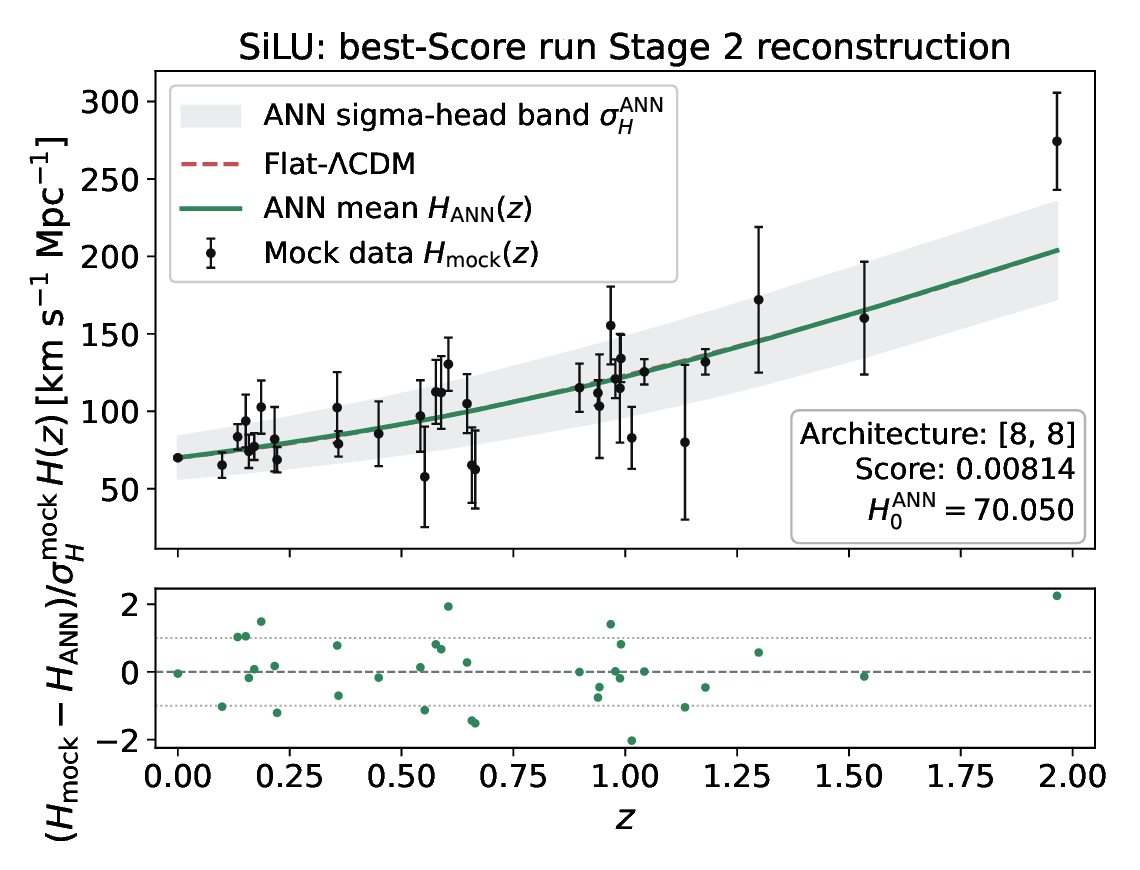}
  }\\
  \subfloat[$\tanh$-$\lbrack64,64\rbrack$: Score $=0.00816$\label{fig:stage2-best-tanh}]{%
    \includegraphics[width=0.492\linewidth]{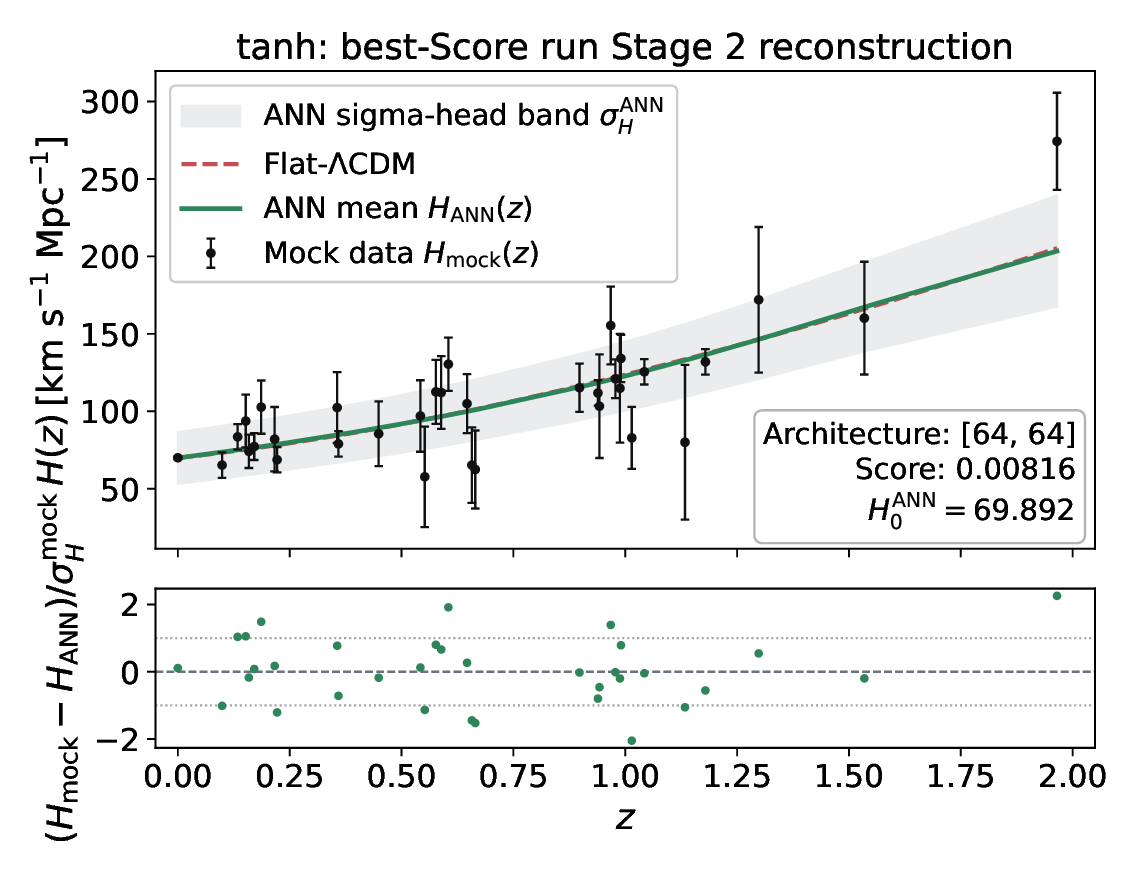}
  }\hfill
  \subfloat[TanhExp-$\lbrack32\rbrack$: Score $=0.00779$\label{fig:stage2-best-tanhexp}]{%
    \includegraphics[width=0.492\linewidth]{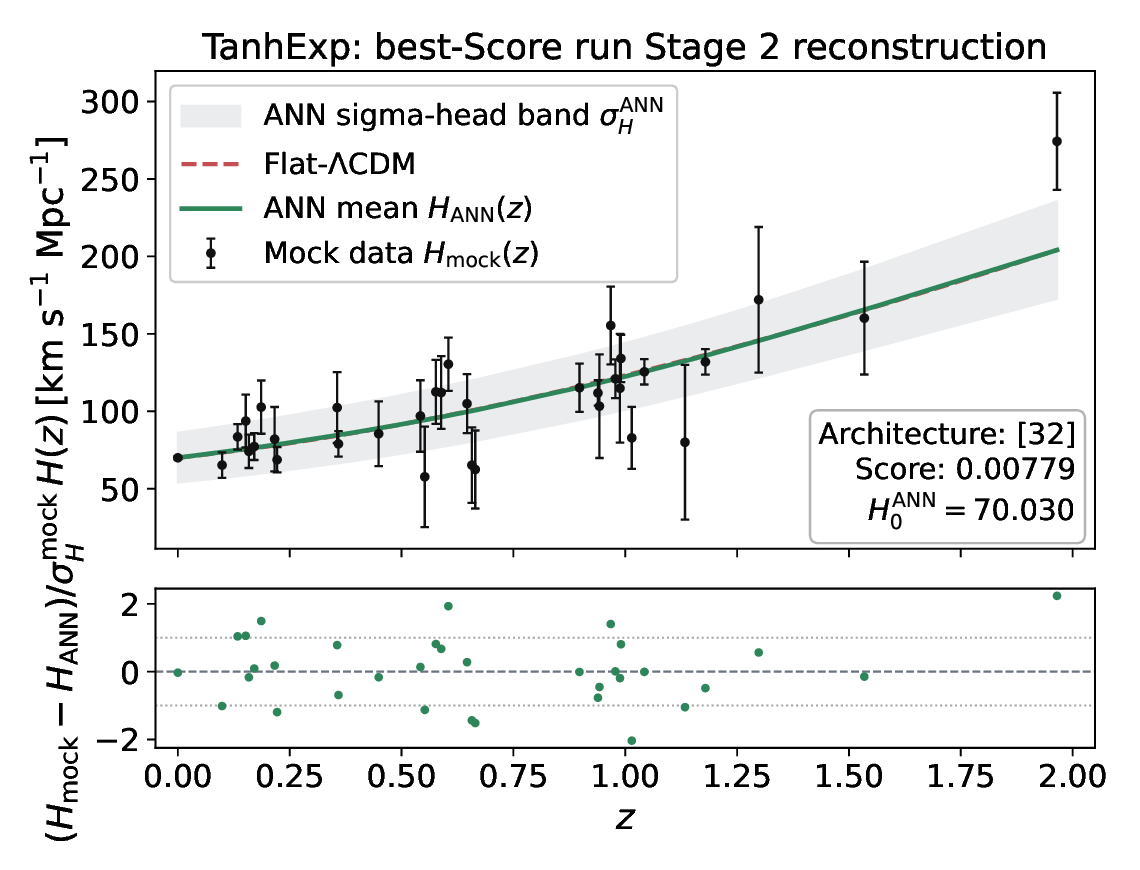}
  }
  \caption{\label{fig:stage2_best_score_examples}
  Best-Score single-run Stage-2 reconstructions selected separately within each activation from its 6,000 full-factorial runs under the same settings, such as weighted L1 loss, zero dropout, learning rate $10^{-3}$, maximum epoch number 800, and no validation split. All four selected examples use mock 3. The Scores are rounded to three significant figures, matching the values printed in the panels. The run metadata are (a) ELU-$[128]$, seed 39, Score $=0.00773$, $H_{0}^{\rm ANN}=69.925$; (b) SiLU-$[8,8]$, seed 65, Score $=0.00814$, $H_{0}^{\rm ANN}=70.050$; (c) $\tanh$-$[64,64]$, seed 94, Score $=0.00816$, $H_{0}^{\rm ANN}=69.892$; and (d) TanhExp-$[32]$, seed 46, Score $=0.00779$, $H_{0}^{\rm ANN}=70.030$, where $H_0$ is in $\mathrm{km\,s^{-1}\,Mpc^{-1}}$. In each upper panel, black points are the 34 mock data, the red dashed curve is the generating flat-$\Lambda$CDM background, the green curve is the single-run ANN reconstruction, and the gray band is the direct scale-head output $\sigma_H^{\rm ANN}$. This single-model scale is distinct from the Stage-3 total ensemble band. The lower panel shows the residual standardized by the mock observational uncertainty, $(H_{\rm mock}-H_{\rm ANN})/\sigma_H^{\rm mock}$; the horizontal lines mark zero and $\pm1$. These best individual runs are illustrative and must not be confused with the candidate-level minimum of $\mathcal{J}$, which selects SiLU-$[8]$.}
\end{figure*}

As illustrated in \zcref{fig:stage2_best_score_examples}, we provide a visual check of the lowest-Score run found separately for each activation function. The selection is made over all 12 structures, five mocks, and 100 seeds for that activation, i.e. 6,000 runs per panel. These examples illustrate individual fits and their scale-head outputs; they are not the aggregate activation--architecture winners defined by $\mathcal{J}$.

Writing the plotted standardized residual as $r_i^{\rm mock}\equiv[H_{\rm mock}(z_i)-H_{\rm ANN}(z_i)]/\sigma_{H,i}^{\rm mock}$, the point-fit diagnostics of the four illustrative runs and the truth-recovery diagnostics of the two leading SiLU candidates are summarized in \zcref{tab:stage2_residual_summary}.

\begin{table}[tbp]
\caption{\label{tab:stage2_residual_summary}
Compact summary of the Stage-2 point-fit and truth-recovery diagnostics. The first three rows give the range across the four illustrative best-Score runs in \zcref{fig:stage2_best_score_examples}. The final two rows use all 500 runs of each candidate in \zcref{tab:stage2_full_factorial}; RMSE is in $\mathrm{km\,s^{-1}\,Mpc^{-1}}$.}
\begin{ruledtabular}
\begin{tabular}{lcc}
Scope & Diagnostic & Value \\
\hline
Best runs & $\langle r^{\rm mock}\rangle$ & $0.019$--$0.034$ \\
Best runs & $\operatorname{RMS}(r^{\rm mock})$ & $0.984$--$0.990$ \\
Best runs & $\max |r^{\rm mock}|$ & $2.24$--$2.34$ \\
SiLU-$[8]$ & $(\sqrt{\overline{S}_{\rm select}},\overline{\mathrm{RMSE}})$ & $(0.266,5.651)$ \\
SiLU-$[16]$ & $(\sqrt{\overline{S}_{\rm select}},\overline{\mathrm{RMSE}})$ & $(0.244,5.078)$ \\
\end{tabular}
\end{ruledtabular}
\end{table}

The similar standardized-residual scales show that none of the four displayed single runs has a large global offset relative to its mock observations. Their interpretation and selection limitations are discussed in \zcref{subsec:disc_activation_loss}.

\subsection{\label{subsec:results_stage3_real}Stage 3: reconstruction of the observed CC compilation}

We next apply the exported Stage-2 configuration to the real CC compilation augmented by one external $H_0$ datum at $z=0$. The adopted configuration is the selected single-hidden-layer SiLU-$[8]$ network with weighted-L1 loss, zero dropout, an active scale head, a learning rate of $10^{-3}$, a maximum of 800 epochs, no validation split, early-stopping patience 80, and gradient clipping at 5.0. For each prior choice, the final prediction is constructed from an explicit ensemble of 100 independently initialized training members.

The Stage-3 execution considers three separate low-redshift anchors: the Planck 2018 prior, $H_0=67.4\pm0.5~\mathrm{km\,s^{-1}\,Mpc^{-1}}$~\cite{Planck:2020Params}; the TRGB prior, $H_0=69.8\pm1.9~\mathrm{km\,s^{-1}\,Mpc^{-1}}$~\cite{Freedman:2019TRGB}; and the SH0ES R21 prior, implemented in the archived configuration as the rounded value $H_0=73.3\pm1.04~\mathrm{km\,s^{-1}\,Mpc^{-1}}$~\cite{Riess:2022SH0ES}. These are exactly the three values used in the executed Stage-3 sweep. The corresponding ANN predictions at $z=0$ are
\begin{align}
H_0^{\rm ANN}(\mathrm{Planck~2018}) &= 67.405,\\
H_0^{\rm ANN}(\mathrm{TRGB}) &= 67.808,\\
H_0^{\rm ANN}(\mathrm{SH0ES~R21}) &= 73.253.
\end{align}
The Planck 2018 and SH0ES R21 reconstructions remain close to the imposed low-redshift anchors, whereas the TRGB run yields a lower ensemble-mean value at $z=0$. The remaining CC data continue to constrain the shape of the late-time background. For comparison, the flat-$\Lambda$CDM fit to the same prior-augmented data yields $(H_0,\Omega_{\mathrm{m},0})=(67.417,0.332)$ for the Planck case, $(69.292,0.301)$ for the TRGB case, and $(72.724,0.251)$ for the SH0ES R21 case.

The prior-dependent Stage-3 summary is collected in \zcref{tab:stage3_prior_summary}. Among the three runs, the TRGB-anchored reconstruction gives the smallest RMSE and MAE with respect to the 34-point prior-augmented observed sample, while the SH0ES-anchored case produces the largest residual level. This ordering describes point-prediction performance on the fitted sample and is not evidence that one external $H_0$ prior is statistically preferred. Using the total operational scale, the in-sample 68\% containment fractions are $31/34=0.912$, $32/34=0.941$, and $30/34=0.882$ for the Planck 2018, TRGB, and SH0ES R21 cases, respectively. The accompanying $O\mathrm{m}(z)$ diagnostics show the same prior-conditioned trend in the flat-$\Lambda$CDM reference levels, shifting from $\Omega_{\mathrm{m},0}\simeq0.332$ to $0.301$ and then to $0.251$, consistent with the comparison fits listed in \zcref{tab:stage3_prior_summary}.

\begin{table*}[tbp]
\caption{\label{tab:stage3_prior_summary}
Stage-3 SiLU-$[8]$ ANN reconstructions for the three external $H_0$ priors. The training sample contains the 33 CC measurements plus one $H_0$ datum at $z=0$, and the quoted ANN result is the saved ensemble-mean prediction at the anchor row from the 100-member run. RMSE and MAE are calculated over all 34 prior-augmented observations. $C_{68}$ is the in-sample containment fraction of \zcref{eq:stage3_in_sample_containment}, evaluated with the total scale that combines the mean scale-head output and ensemble dispersion. The three external anchors are the Planck 2018 prior~\cite{Planck:2020Params}, the TRGB prior~\cite{Freedman:2019TRGB}, and the SH0ES R21 prior~\cite{Riess:2022SH0ES}; the SH0ES central value is rounded to one decimal place in the executed configuration. The flat-$\Lambda$CDM reference values are obtained by fitting the same prior-augmented dataset used by the ANN in each panel.}
\begin{ruledtabular}
\begin{tabular}{lccccccc}
Prior
  & $H_0^{\rm prior}$
  & $H_0^{\rm ANN}$
  & $H_0^{\Lambda{\rm CDM}}$
  & $\Omega_{\mathrm{m},0}^{\Lambda{\rm CDM}}$
  & $\mathrm{RMSE}_{\rm obs-pred}$
  & $\mathrm{MAE}_{\rm obs-pred}$
  & $C_{68}$ \\
\hline
Planck 2018 & $67.4\pm0.50$ & 67.405 & 67.417 & 0.332 & 12.453 & 8.547 & 0.912 \\
TRGB & $69.8\pm1.90$ & 67.808 & 69.292 & 0.301 & 11.498 & 8.244 & 0.941 \\
SH0ES R21 & $73.3\pm1.04$ & 73.253 & 72.724 & 0.251 & 13.324 & 9.688 & 0.882 \\
\end{tabular}
\end{ruledtabular}
\end{table*}

\begin{figure*}[tbp]
  \centering
  \subfloat[$H(z)$ reconstruction and residual standardized by $\sigma_{\rm data}$ for the Planck 2018 prior\label{fig:stage3-p18-a}]{%
    \includegraphics[width=0.492\linewidth]{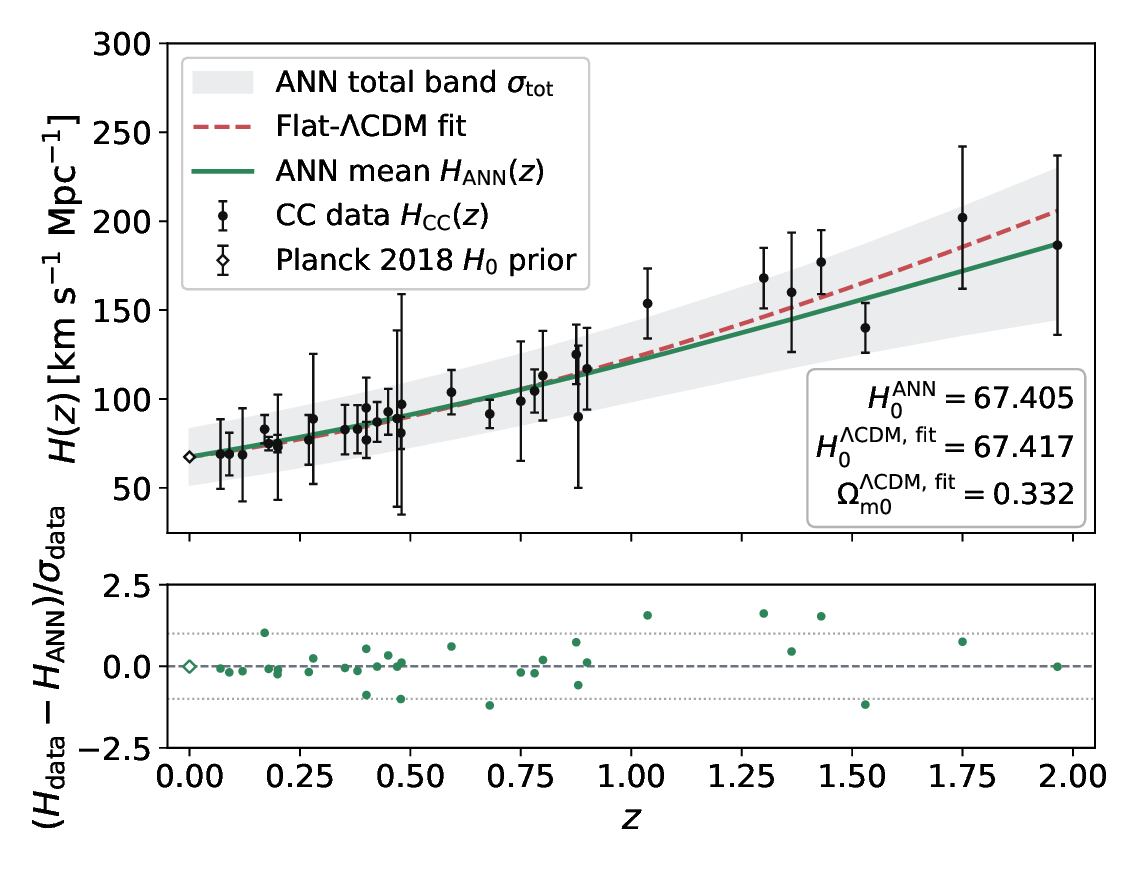}
  }\hfill
  \subfloat[$O\mathrm{m}(z)$ diagnostic for the Planck 2018 prior\label{fig:stage3_p18-b}]{%
    \includegraphics[width=0.492\linewidth]{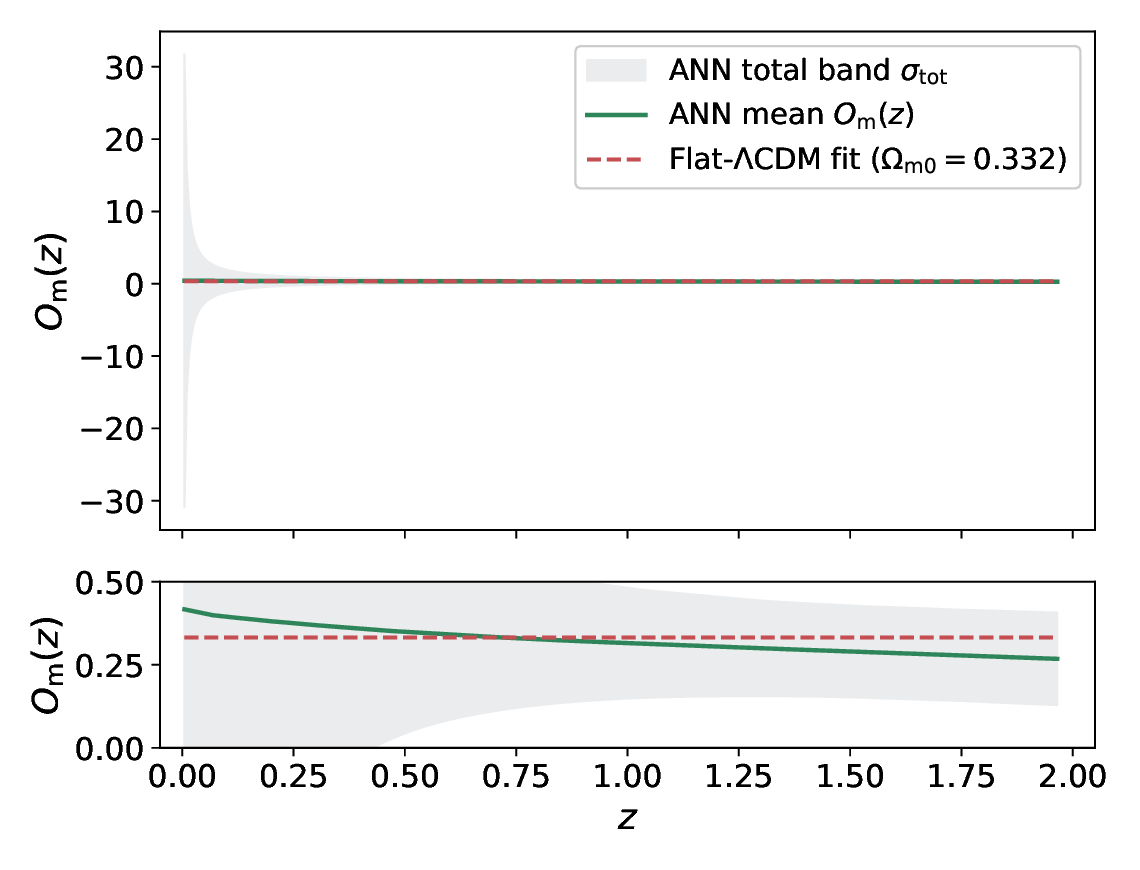}
  }
  \caption{\label{fig:stage3_p18}
  Stage-3 SiLU-$[8]$ reconstruction for the Planck 2018 anchor, $H_0=67.4\pm0.5~\mathrm{km\,s^{-1}\,Mpc^{-1}}$~\cite{Planck:2020Params}. In panel (a), the upper part shows the 33 CC measurements, the $z=0$ prior point, the ANN ensemble mean with its total operational band $\sigma_{\rm tot}$, and the flat-$\Lambda$CDM fit to the same prior-augmented dataset. The inset reports $H_0^{\rm ANN}=67.405$, $H_0^{\Lambda{\rm CDM}}=67.417$, and $\Omega_{\mathrm{m},0}^{\Lambda{\rm CDM}}=0.332$. The lower part shows the residual standardized by the reported observational uncertainty, $(H_{\rm data}-H_{\rm ANN})/\sigma_{\rm data}$, with dashed zero and dotted $\pm1$ reference lines. In panel (b), the upper part shows $O\mathrm{m}(z)$ normalized by the ANN value $H_0^{\rm ANN}$, and the lower part shows the same diagnostic over $0\le O\mathrm{m}(z)\le0.5$. The shaded $O\mathrm{m}(z)$ band is propagated from $\sigma_{\rm tot}$ using \zcref{eq:om_uncertainty_stage3}; the dashed horizontal line is the flat-$\Lambda$CDM reference $\Omega_{\mathrm{m},0}=0.332$.}
\end{figure*}

\begin{figure*}[tbp]
  \centering
  \subfloat[$H(z)$ reconstruction and residual standardized by $\sigma_{\rm data}$ for the TRGB prior\label{fig:stage3-trgb-a}]{%
    \includegraphics[width=0.492\linewidth]{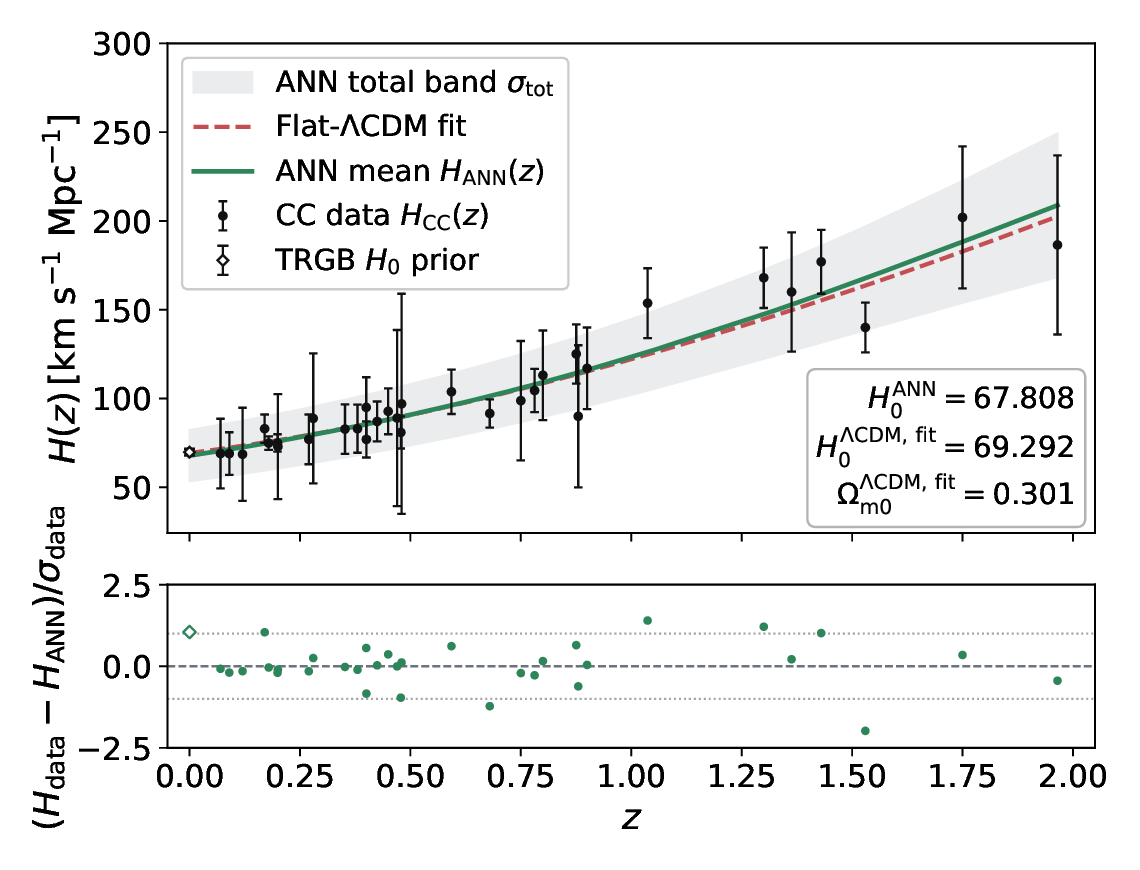}
  }\hfill
  \subfloat[$O\mathrm{m}(z)$ diagnostic for the TRGB prior\label{fig:stage3-trgb-b}]{%
    \includegraphics[width=0.492\linewidth]{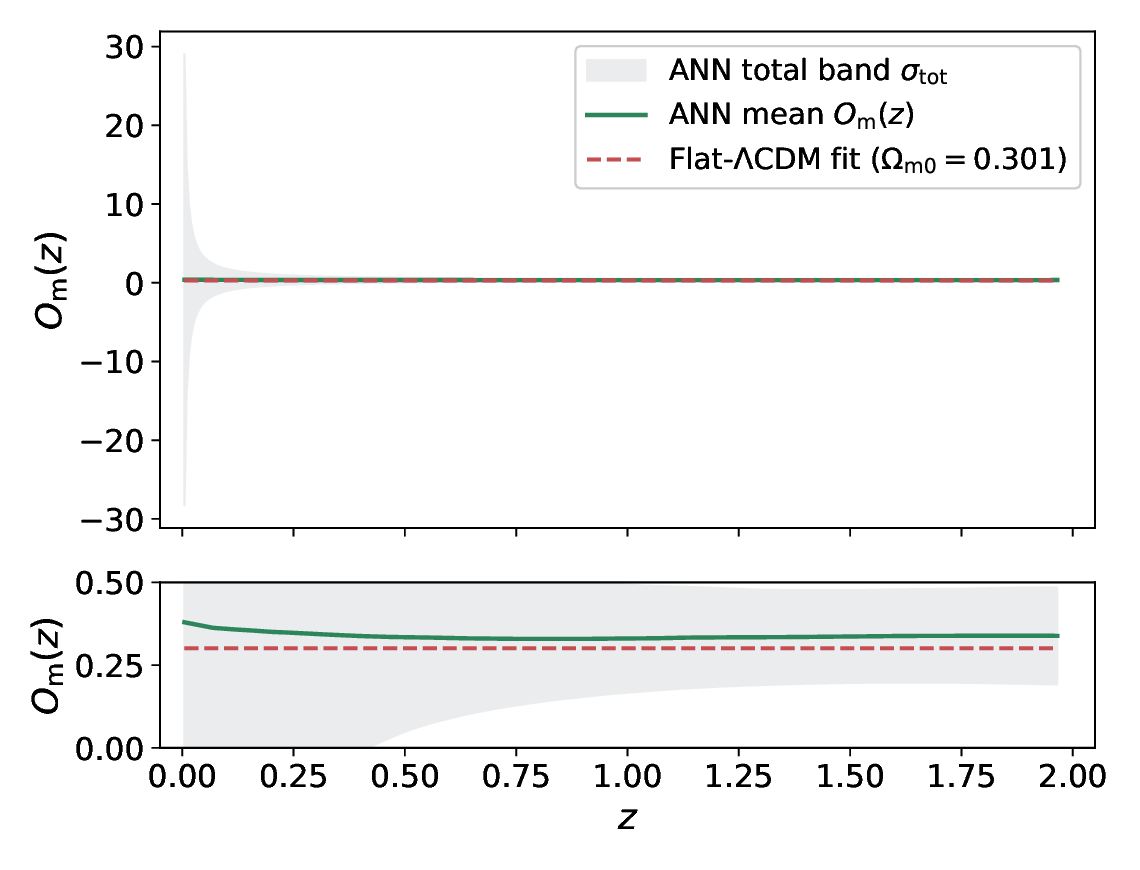}
  }
  \caption{\label{fig:stage3_trgb}
  Stage-3 SiLU-$[8]$ reconstruction for the TRGB anchor, $H_0=69.8\pm1.9~\mathrm{km\,s^{-1}\,Mpc^{-1}}$~\cite{Freedman:2019TRGB}. Panels (a) and (b) use the same layout, total operational band, standardization by $\sigma_{\rm data}$, and $O\mathrm{m}(z)$ conventions as \zcref{fig:stage3_p18}. In panel (a), the inset values are $H_0^{\rm ANN}=67.808$, $H_0^{\Lambda{\rm CDM}}=69.292$, and $\Omega_{\mathrm{m},0}^{\Lambda{\rm CDM}}=0.301$. The ANN prediction at $z=0$ is $1.992~\mathrm{km\,s^{-1}\,Mpc^{-1}}$ below the TRGB central value. In panel (b), the dashed horizontal line marks the flat-$\Lambda$CDM reference $\Omega_{\mathrm{m},0}=0.301$.}
\end{figure*}

\begin{figure*}[tbp]
  \centering
  \subfloat[$H(z)$ reconstruction and residual standardized by $\sigma_{\rm data}$ for the SH0ES R21 prior\label{fig:stage3-r21-a}]{%
    \includegraphics[width=0.492\linewidth]{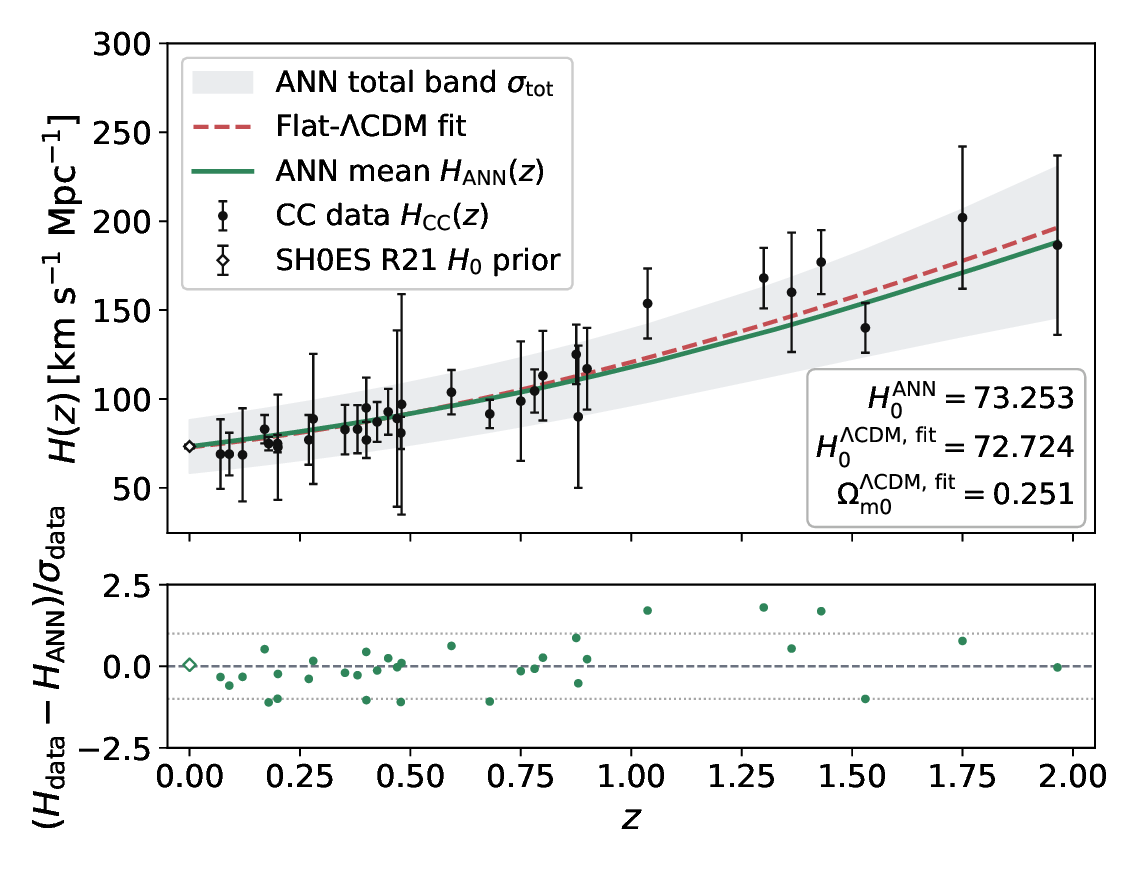}
  }\hfill
  \subfloat[$O\mathrm{m}(z)$ diagnostic for the SH0ES R21 prior\label{fig:stage3-r21-b}]{%
    \includegraphics[width=0.492\linewidth]{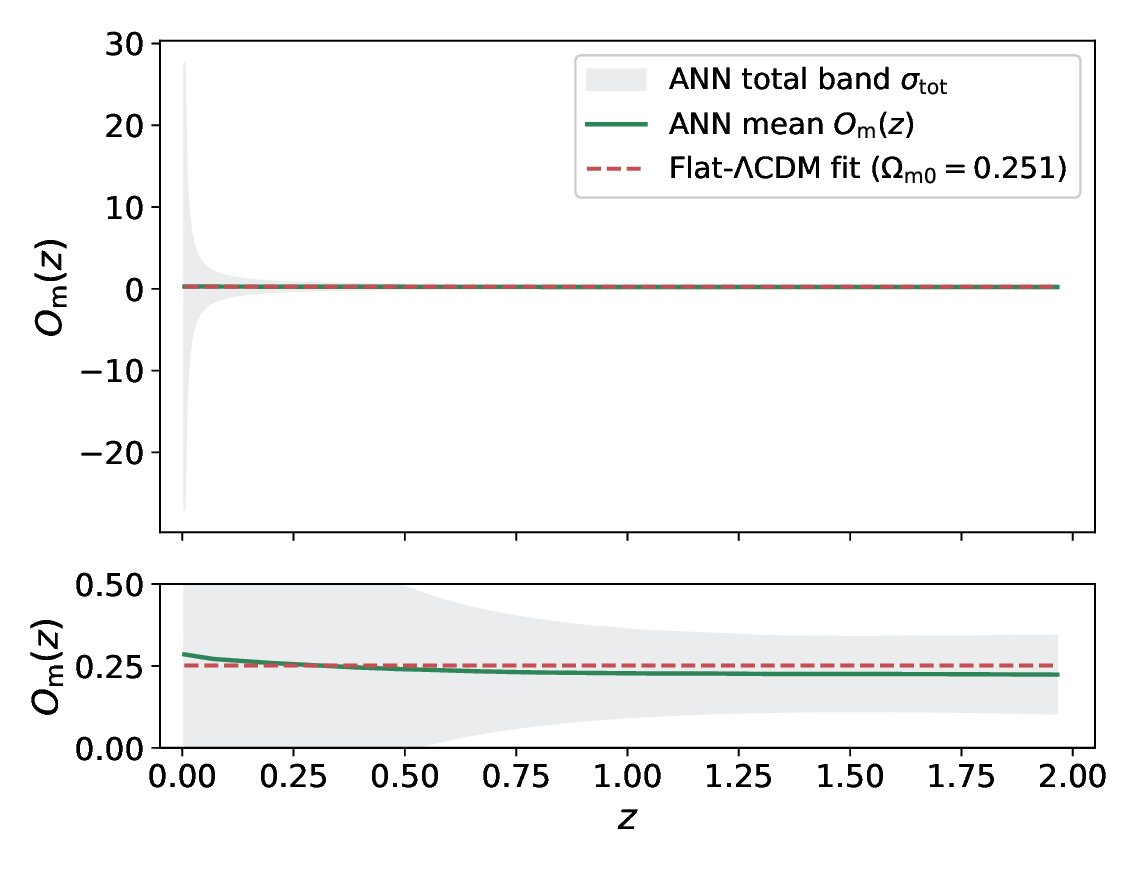}
  }
  \caption{\label{fig:stage3_r21}
  Stage-3 SiLU-$[8]$ reconstruction for the SH0ES R21 anchor, implemented as $H_0=73.3\pm1.04~\mathrm{km\,s^{-1}\,Mpc^{-1}}$~\cite{Riess:2022SH0ES}. Panels (a) and (b) use the same layout, total operational band, standardization by $\sigma_{\rm data}$, and $O\mathrm{m}(z)$ conventions as \zcref{fig:stage3_p18}. In panel (a), the inset values are $H_0^{\rm ANN}=73.253$, $H_0^{\Lambda{\rm CDM}}=72.724$, and $\Omega_{\mathrm{m},0}^{\Lambda{\rm CDM}}=0.251$. This case has the largest RMSE and MAE of the three runs. In panel (b), the dashed horizontal line marks the correspondingly lowest flat-$\Lambda$CDM reference, $\Omega_{\mathrm{m},0}=0.251$.}
\end{figure*}

The prior-conditioned Stage-3 results are organized into one figure for each external anchor in \zcref{fig:stage3_p18,fig:stage3_trgb,fig:stage3_r21}. In all cases, panel (a) in \zcref{fig:stage3-p18-a,fig:stage3-trgb-a,fig:stage3-r21-a} contains the Hubble reconstruction and the residuals standardized by the reported observational uncertainty, $r_i^{\mathrm{obs},(k)}=[H_i-\bar H^{(k)}(z_i)]/\sigma_{\mathrm{data},i}$, while panel (b) in \zcref{fig:stage3_p18-b,fig:stage3-trgb-b,fig:stage3-r21-b} shows the corresponding $O\mathrm{m}(z)$ diagnostic. In panel (b), the upper part shows the full diagnostic range and the lower part shows the same curve and propagated band with the vertical range restricted to $0\le O\mathrm{m}(z)\le0.5$ for comparison with the flat-$\Lambda$CDM matter-density reference. The exact $z=0$ point, where the $O\mathrm{m}(z)$ denominator vanishes, is omitted. The SiLU-$[8]$ architecture and all training settings are otherwise identical across the three runs.

A complementary check is provided by the lower parts of panels (a) in \zcref{fig:stage3-p18-a,fig:stage3-trgb-a,fig:stage3-r21-a}, using residuals that have been standardized by the reported observational uncertainties. For the Planck 2018, TRGB, and SH0ES R21 runs, respectively, the mean values of $r_i^{\mathrm{obs},(k)}$ are $0.097$, $0.043$, and $0.012$, while their root-mean-square values are $0.683$, $0.687$, and $0.759$. Each case has seven of the 34 points with $|r_i^{\mathrm{obs},(k)}|>1$, and the largest absolute standardized residuals lie between $1.62$ and $1.98$. The near-zero means show no large signed offset in the fitted sample, whereas the somewhat larger root-mean-square value for the SH0ES R21 case is consistent with its larger RMSE and MAE. These are in-sample residual diagnostics and do not by themselves measure generalization. They also differ from $C_{68}$: the plotted residuals are standardized by $\sigma_{\rm data}$, whereas the containment statistic and the standardized residuals of \zcref{eq:norm_resid_stage3_real} use the ANN total operational scale $\sigma_{\rm tot}$.

\section{\label{sec:discussion}Discussions}

In this section, we interpret the Stage-2 and Stage-3 results, assess their methodological implications, and summarize the main limitations of the current baseline implementation.

\subsection{\label{subsec:disc_activation_loss}Conditional interpretation and limitations of the Stage-2 design search}

The Stage-2 ranking reported in \zcref{subsec:results_stage2} should be interpreted as the selection of an activation--architecture configuration under the adopted mock-generation, training, and aggregation protocol, rather than as the identification of a universally optimal ANN model. This distinction is evident from the two aggregation rules: the primary hierarchical statistic $\mathcal{J}$ of \zcref{eq:stage2_activation_statistic} selects SiLU-$[8]$, whereas the pooled-median rule of \zcref{eq:stage2_architecture_export_rule} selects SiLU-$[16]$. Moreover, \zcref{tab:stage2_full_factorial} shows that SiLU-$[16]$ has smaller pooled mean truth-recovery error and RMSE. Thus, the term \emph{best} refers specifically to the minimum of the predefined hierarchical statistic within the candidate grid of \zcref{eq:stage2_full_factorial_grid}; it does not imply simultaneous superiority in every diagnostic. Although the preference for the SiLU activation is retained under this particular change of aggregation, the preferred width is not invariant.

The robustness of this selection is further limited by both the finite mock suite and the fixed scope of the search. The 100 initializations per mock provide a detailed characterization of seed-to-seed variation for a fixed candidate and dataset, while the within-mock spread and the separation among the mock-specific medians in \zcref{fig:stage2_score_depth1,fig:stage2_score_depth2} indicate initialization sensitivity and realization sensitivity, respectively. However, only five mock datasets enter $\mathcal{J}$, and all five share the fiducial expansion history and generation procedure specified in \zcref{subsec:mock_generation_cc}. They therefore probe differences in sampled redshifts, uncertainties, and noise realizations, but not variation across independent cosmological-background families. No independent mock suite is retained for post-selection evaluation under the training protocol of \zcref{subsec:training_selection_cc}. In addition, only the activation function and architecture are varied; the weighted-L1 loss, smoothness weight, optimizer, learning rate, dropout, early-stopping prescription, scale-head configuration, and data-generation procedure remain fixed. Consequently, the higher and more dispersed Scores of several wider two-hidden-layer candidates support a compact baseline within the present grid, but do not establish a general preference for shallow networks outside this experimental design.

A particularly important conditioning arises from the background model used to generate the mocks. All five Stage-2 datasets are noisy realizations of the same smooth fiducial flat-$\Lambda$CDM expansion history defined in \zcref{eq:fiducial_values_stage1,eq:Hz_flat_lcdm}. This background enters the selection directly through $S_{\mathrm{select}}$ in \zcref{eq:s_select}, which evaluates recovery of the known mock truth. Its influence may also interact with the smoothness term in \zcref{eq:r_smooth,eq:architecture_score}, because the current mock family contains neither localized features nor alternative redshift evolution that might favor a different model capacity. The simplicity and smoothness of the generated datasets may therefore contribute to the selection of a compact architecture. Nevertheless, flat-$\Lambda$CDM is not imposed as the functional form of the Stage-3 reconstruction: once the configuration has been selected, Stage~3 reconstructs the observed CC data non-parametrically within the function class represented by the network. Since the full design search has not been repeated with alternative background cosmologies, the present analysis neither demonstrates that the selected configuration would change nor establishes its complete independence from the assumed mock background. The magnitude and direction of this possible sensitivity remain unquantified.

Within these constraints, the standardized residuals provide a complementary assessment of how the selected runs reproduce the mock observations. Denoting the injected fluctuation and truth-standardized reconstruction error by $\eta_i$ and $r_i^{\rm truth}$, respectively, the residual displayed in \zcref{fig:stage2_best_score_examples} can be written as
\begin{equation}
\begin{aligned}
r_i^{\rm mock}
&\equiv
\frac{H_{\rm mock}(z_i)-H_{\rm ANN}(z_i)}{\sigma_{H,i}^{\rm mock}}
=\eta_i-r_i^{\rm truth},
\\
\eta_i
&\equiv
\frac{H_{\rm mock}(z_i)-H_{\rm true}(z_i)}{\sigma_{H,i}^{\rm mock}}
\sim\mathcal{N}(0,1),
\\
r_i^{\rm truth}
&\equiv
\frac{H_{\rm ANN}(z_i)-H_{\rm true}(z_i)}{\sigma_{H,i}^{\rm mock}}.
\end{aligned}
\label{eq:stage2_residual_decomposition}
\end{equation}

Under this decomposition, the weighted-L1 term in \zcref{eq:weighted_l1_cc} measures agreement with a noisy mock dataset, whereas $S_{\mathrm{select}}$ evaluates recovery of its generating background. As summarized in \zcref{tab:stage2_residual_summary}, the four displayed runs have residual means close to zero, RMS values close to unity, and maximum absolute residuals of approximately $2.3$. This behavior is consistent with reconstructions that follow the smooth background while leaving approximately the injected noise in the residuals; exact interpolation would instead drive $\operatorname{RMS}(r^{\rm mock})$ toward zero. At candidate level, $\sqrt{\overline{S}_{\rm select}}=0.266$ for SiLU-$[8]$ indicates a 500-run truth-recovery error of approximately $0.27$ times the mock observational scale. Nevertheless, these diagnostics remain descriptive and in sample. In particular, the four panels in \zcref{fig:stage2_best_score_examples} are extreme single-run selections from 6,000 runs per activation, and all correspond to mock~3. Their similar residual scales therefore do not establish equivalent activation performance, absence of overfitting, or generalization to independent mock families. Likewise, a near-zero global residual mean does not exclude a sign-changing redshift dependence.

Taken together, these considerations support SiLU-$[8]$ as a reproducible baseline for the present mock and training protocol, but not as a universal optimum. A direct robustness study should repeat the complete Stage-2 search across alternative mock backgrounds while holding the redshift and noise prescriptions fixed, and should separately vary the sampling and uncertainty models at a fixed background. Comparing the resulting candidate rankings, selection frequencies, and Stage-3 reconstructions would distinguish background-model sensitivity from sampling and noise sensitivity.

\subsection{\label{subsec:disc_stage3_interpretation}Interpretation of the Stage-3 CC reconstruction}

The Stage-3 reconstructions presented in \zcref{subsec:results_stage3_real,fig:stage3_p18,fig:stage3_trgb,fig:stage3_r21} should be read as prior-conditioned fits to the common observed CC compilation, rather than as prior-free determinations of $H_0$. The Planck 2018 and SH0ES R21 ensemble means closely follow their respective low-redshift anchors, whereas the TRGB case shows the larger endpoint displacement summarized in \zcref{tab:stage3_prior_summary}. Because the architecture, training settings, and 33 CC measurements are common to all three runs, this contrast motivates a closer examination of how the external $H_0$ uncertainty enters the weighted-L1 loss function.

To examine the TRGB behavior more closely, the normalization of \zcref{eq:h_normalization} gives $h_{\mathrm{scale}}=90.8~\mathrm{km\,s^{-1}\,Mpc^{-1}}$ for all three prior-augmented observed samples. Under the weighted-L1 loss of \zcref{eq:weighted_l1_cc}, the coefficient multiplying the absolute residual of the $H_0$ anchor for prior choice $k$ and the corresponding relative weights are
\begin{equation}
\begin{aligned}
w_{H_0}^{(k)}
=
\frac{1}{s_{H_0}^{(k)}+\varepsilon_s}
=
\frac{1}{\sigma_{H_0}^{(k)}/h_{\mathrm{scale}}+\varepsilon_s},
\\
w_{H_0}^{\mathrm{P18}}
:
w_{H_0}^{\mathrm{TRGB}}
:
w_{H_0}^{\mathrm{R21}}
\simeq
3.80:1:1.83.
\end{aligned}
\label{eq:h0_prior_l1_weight}
\end{equation}
The ratio in \zcref{eq:h0_prior_l1_weight}, calculated from the uncertainties $0.50$, $1.90$, and $1.04~\mathrm{km\,s^{-1}\,Mpc^{-1}}$ while neglecting the small numerical floor, implies that the TRGB anchor has approximately $26\%$ of the Planck weight and $55\%$ of the SH0ES R21 weight for the same absolute normalized prediction error. Its broader uncertainty may therefore constrain the $z=0$ endpoint less strongly, allowing the continuation of the curve constrained by the 33 CC measurements to exert greater influence on $H_0^{\rm ANN}$. The displacement $H_0^{\rm ANN}-H_0^{\rm prior}=-1.992~\mathrm{km\,s^{-1}\,Mpc^{-1}}$ corresponds to $-1.05$ times the TRGB uncertainty and is not in strong tension with that prior. Nevertheless, this weighting argument does not establish a unique cause: the prior central values and uncertainties change simultaneously across the three runs, and the flat-$\Lambda$CDM fit to the same TRGB-augmented sample gives $H_0^{\Lambda{\rm CDM}}=69.292~\mathrm{km\,s^{-1}\,Mpc^{-1}}$. The broader TRGB uncertainty is therefore a plausible contributor, but not the sole demonstrated explanation for the lower ANN endpoint.

Turning from the endpoint behavior to fitted-sample performance, the RMSE and MAE values in \zcref{tab:stage3_prior_summary} provide complementary measures of point-prediction error: RMSE gives greater weight to large residuals, whereas MAE is less sensitive to individual extremes. Their common ordering shows that the larger residual level in the SH0ES R21 run is not confined to the squared-error metric. Even so, the three reconstructions share the same 33 CC measurements, no sampling uncertainty is reported for the metric differences, and no held-out sample is used. The ordering is therefore descriptive and cannot establish either a statistically significant preference among the external priors or an out-of-sample performance ranking. For a direct comparison of the ANN and flat-$\Lambda$CDM reconstructions on the same observations, we use the raw observational chi-square statistic
\begin{equation}
\chi_X^2(\mathcal{S})
=
\sum_{i\in\mathcal{S}}
\left[
\frac{H_X(z_i)-H_{\mathrm{data}}(z_i)}{\sigma_{\mathrm{data},i}}
\right]^2,
\quad
X\in\{\mathrm{ANN},\Lambda\mathrm{CDM}\},
\label{eq:stage3_ann_lcdm_chi2}
\end{equation}
Here, $\mathcal{S}_{H_0+\mathrm{CC}}$ contains the 33 CC measurements and the $z=0$ prior point, whereas $\mathcal{S}_{\mathrm{CC}}$ excludes only that anchor. The flat-$\Lambda$CDM parameters are fitted once to $\mathcal{S}_{H_0+\mathrm{CC}}$, as in the Stage-3 figures; the CC-only column in \zcref{tab:stage3_ann_lcdm_chi2} evaluates the same fitted curve after removing the anchor contribution rather than refitting it. The ANN columns are evaluated analogously from the same prior-conditioned reconstruction.

\begin{table*}[tbp]
\caption{\label{tab:stage3_ann_lcdm_chi2}
Raw in-sample observational chi-square statistics from \zcref{eq:stage3_ann_lcdm_chi2}. The $H_0+\mathrm{CC}$ scope contains 34 points and the CC-only scope contains 33 points. These values use the reported observational uncertainties and include no parameter-count subtraction or complexity penalty.}
\begin{ruledtabular}
\begin{tabular}{lcccc}
Prior
  & \multicolumn{2}{c}{$H_0+\mathrm{CC}$ (34 points)}
  & \multicolumn{2}{c}{$\mathrm{CC}$ only (33 points)} \\
  & $\chi_{\mathrm{ANN}}^2$
  & $\chi_{\Lambda\mathrm{CDM}}^2$
  & $\chi_{\mathrm{ANN}}^2$
  & $\chi_{\Lambda\mathrm{CDM}}^2$ \\
\hline
Planck 2018 & 15.840 & 14.813 & 15.840 & 14.812 \\
TRGB & 16.051 & 15.000 & 14.952 & 14.929 \\
SH0ES R21 & 19.597 & 17.542 & 19.595 & 17.235 \\
\end{tabular}
\end{ruledtabular}
\end{table*}

In this comparison, the flat-$\Lambda$CDM values for the 34-point scope are lower than the corresponding ANN values by $1.027$, $1.051$, and $2.055$ for Planck 2018, TRGB, and SH0ES R21, respectively. The subset values in \zcref{tab:stage3_ann_lcdm_chi2} further show that the TRGB CC-only difference is only $0.023$, so most of its 34-point difference is associated with agreement at the $H_0$ anchor; this behavior is consistent with the weaker weighted-L1 influence discussed above. By contrast, the Planck 2018 and SH0ES R21 CC-only differences of $1.028$ and $2.360$ arise mainly from the common CC measurements. The identical displayed Planck ANN values for the two scopes result from rounding because the anchor contribution is negligible. Importantly, these raw statistics do not define a likelihood-ratio or model-selection test: the ANN is trained with the weighted-L1 objective, the two-parameter flat-$\Lambda$CDM reference minimizes observational chi-square, and no effective ANN parameter count is assigned. Consequently, the comparison shows descriptively that the ANN does not attain a lower observational chi-square, while the operational bands in \zcref{fig:stage3_p18,fig:stage3_trgb,fig:stage3_r21} reveal no resolved departure from the corresponding prior-conditioned flat-$\Lambda$CDM curves; neither result establishes a statistical preference between the two reconstruction approaches.

A complementary perspective is provided by the $O\mathrm{m}(z)$ diagnostic defined in \zcref{eq:om_diagnostic_stage3}. The curves in \zcref{fig:stage3_p18,fig:stage3_trgb,fig:stage3_r21} show prior-conditioned shape differences: the Planck 2018 and SH0ES R21 cases cross their respective flat-$\Lambda$CDM reference levels, whereas the TRGB curve remains above its reference throughout the sampled CC range. Their RMS deviations from those levels are $0.0366$, $0.0393$, and $0.0173$, respectively. Despite these trends, every reference level remains inside the propagated operational $1\sigma$ band at all CC redshifts, with maximum absolute deviations standardized by that band of $0.462$, $0.265$, and $0.231$. The trends are therefore not resolved departures under the present uncertainty prescription. Moreover, proximity to the horizontal reference and goodness of fit to the CC measurements are not interchangeable: SH0ES R21 has the smallest RMS $O\mathrm{m}(z)$ deviation but the largest ANN CC-only chi-square, while the positive TRGB offset accompanies nearly identical ANN and flat-$\Lambda$CDM CC-only values. Accordingly, $O\mathrm{m}(z)$ is used here to characterize prior-conditioned shape differences, not to rank the external $H_0$ priors by data-fit quality.

Finally, the operational uncertainty band requires a distinct interpretation. As defined in \zcref{eq:sigma_total_cc}, it combines the mean scale-head output with the dispersion across 100 independently initialized ensemble members. The averages in \zcref{tab:stage3_uncertainty_diagnostics} show that the mean scale-head output is substantially larger than the ensemble dispersion and is numerically close to the mean reported data uncertainty. The ensemble component therefore characterizes initialization variation for the fixed dataset, architecture, loss, and training prescription, but does not exhaust epistemic or systematic uncertainty.

\begin{table*}[tbp]
\caption{\label{tab:stage3_uncertainty_diagnostics}
Stage-3 in-sample uncertainty diagnostics. Angle brackets denote averages over the 34 prior-augmented training points; all scale columns are in $\mathrm{km\,s^{-1}\,Mpc^{-1}}$. The last column is the standard deviation of the residual in \zcref{eq:norm_resid_stage3_real}, computed over the 34 fitted points with divisor $N$.}
\begin{ruledtabular}
\begin{tabular}{lcccccc}
Prior
  & $C_{68}$
  & $\langle\sigma_{\rm data}\rangle$
  & $\langle\bar\sigma_{\rm head}\rangle$
  & $\langle\sigma_{\rm ens}\rangle$
  & $\langle\sigma_{\rm tot}\rangle$
  & $\operatorname{SD}(r^{\rm norm})$ \\
\hline
Planck 2018 & $31/34$ & 21.133 & 19.507 & 4.068 & 20.402 & 0.509 \\
TRGB & $32/34$ & 21.174 & 18.916 & 2.816 & 19.397 & 0.502 \\
SH0ES R21 & $30/34$ & 21.149 & 19.071 & 3.782 & 19.785 & 0.575 \\
\end{tabular}
\end{ruledtabular}
\end{table*}

Even so, the containment values cannot be interpreted as calibrated predictive coverage. One observation changes $C_{68}$ by $1/34=0.029$, so the three counts differ by at most two points; moreover, the same 34 prior-augmented observations are used for both training and evaluation, with no held-out Stage-3 sample. The values $\operatorname{SD}(r^{\rm norm})=0.502$--$0.575$ therefore establish only that the fitted-sample residuals are narrower than the operational band. Although the large in-sample containment fractions are consistent with broad bands on the fitted points, they do not demonstrate conservative coverage for the latent expansion history or for new observations. The band should consequently be treated as an operational uncertainty summary rather than as a Bayesian posterior predictive interval or a calibrated frequentist interval.

\subsection{\label{subsec:disc_limitations}Current limitations and future extensions}

Several limitations follow directly from the archived execution analyzed here. First, we treat the observed CC compilation with one external $H_0$ anchor at a time, rather than performing a joint covariance-aware inference from CC, Type Ia supernovae, and BAO. This is appropriate for establishing a transparent baseline, but we have not determined how the same staged design logic behaves for multi-probe likelihoods. Second, as discussed in \zcref{subsec:disc_activation_loss}, all five mock datasets share the smooth fiducial flat-$\Lambda$CDM background and generation procedure. Their redshift sampling, uncertainties, noise, and functional complexity enter the Score-based comparison, so we have not quantified the sensitivity of the selected configuration to alternative backgrounds or localized features. Finally, we use no Stage-3 validation split or independent calibration sample, and the ensemble dispersion varies only the initialization while holding the dataset and modeling prescription fixed. As a result, we treat the band assessed in \zcref{subsec:disc_stage3_interpretation} as an operational uncertainty summary rather than as a calibrated frequentist interval or a complete Bayesian posterior predictive interval.

These constraints motivate complementary extensions. We will first repeat the complete Stage-2 search across alternative background histories while varying the sampling and uncertainty prescriptions independently, and then propagate the same staged design logic to multi-probe forward models involving Pantheon+ and BAO observables. In parallel, we will compare the combined scale-head and ensemble prescription with covariance-aware objectives and alternative uncertainty parametrizations using independent calibration data. Finally, we will broaden the present $O\mathrm{m}(z)$ analysis into a jointly validated set of derived diagnostics, including $E(z)$ and numerical derivatives of $H(z)$, for null tests and model comparison.

\section{\label{sec:conclusion}Conclusions}

In this manuscript, we have proposed an ANN framework for reconstructing the late-time expansion history from the observed CC compilation. As summarized in \zcref{subsec:stage_pipeline_cc}, the baseline separates CC-like mock generation, Stage-2 activation--architecture selection, and Stage-3 reconstruction of the observed CC compilation augmented by one external $H_0$ prior at $z=0$.

In Stage~2, we have evaluated all 48 combinations of four activation functions and 12 architectures at fixed weighted-L1 loss, as reported in \zcref{subsec:results_stage2}. The full-factorial comparison has shown that the relative performance of the tested activation functions can vary with network architecture, so the two choices should not be treated as independent. Moreover, because only four activation functions have been examined, the observed performance trends cannot be assumed to extend to other activation functions not included in our baseline. Using the predefined primary hierarchical rule, we have identified the compact SiLU-$[8]$ candidate with $\mathcal{J}=0.0473$, whereas the pooled-median sensitivity check has favored SiLU-$[16]$; accordingly, the primary-rule output has been used for Stage~3. The exported configuration is therefore regarded as conditional on the adopted aggregation rule, candidate grid, and mock-generation protocol. Although the Stage-3 reconstruction does not impose the fiducial flat-$\Lambda$CDM functional form, all five mock datasets used for selection share this background. Accordingly, neither the complete independence of the selected configuration from the assumed mock background nor the extent to which that background influences the subsequent reconstruction has been established, as discussed in \zcref{subsec:disc_activation_loss}.

Using the Stage-2 selected ANN architecture, we have produced three prior-conditioned 100-member ensemble reconstructions, which are presented in \zcref{subsec:results_stage3_real}. The summary in \zcref{tab:stage3_prior_summary} shows that the ensemble endpoint responds differently to the three anchors, with the largest displacement occurring for TRGB. Across the three runs, the observed-space RMSE values range from $11.498$ to $13.324~\mathrm{km\,s^{-1}\,Mpc^{-1}}$, and the in-sample 68\% containment fractions range from $0.882$ to $0.941$. Because these quantities have been evaluated on the fitted prior-augmented samples, we do not interpret the containment fractions as calibrated predictive coverage; instead, the reported band is used as the operational uncertainty summary discussed in \zcref{subsec:disc_stage3_interpretation}.

Taken together, we have established a reproducible staged baseline in which the ANN design transferred to the observed CC sample has been fixed through a documented mock-based selection procedure. Rather than identifying a universal architecture or providing a prior-free reconstruction, the framework makes the conditioning assumptions and selection rules explicit. This separation improves the auditability of ANN-based cosmological reconstruction and provides a controlled basis for subsequent robustness studies.

In future work, we will pursue the extensions summarized in \zcref{subsec:disc_limitations} by repeating the full Stage-2 search with alternative mock backgrounds and independently varied sampling and uncertainty prescriptions, extending the framework to covariance-aware multi-probe analyses, validating alternative uncertainty parametrizations on independent data, and broadening the derived-diagnostic analysis beyond the present $O\mathrm{m}(z)$ implementation.

\begin{acknowledgments}
The work of Yuki Hashimoto was supported by JST SPRING, Japan Grant Number JPMJSP2190.
The work of Kazuharu Bamba was also supported in part by the JSPS KAKENHI Grant Number 25KF0176 and a grant-in-aid of academic research of Yamaguchi Scholarship Foundation. Furthermore, this work was supported from the Research Promotion Project of Fukushima University Fund.
\end{acknowledgments}

\appendix

\section{\label{append:mock_datasets}Mock generation details}
\begin{figure*}[p]
  \centering

  \captionsetup[subfloat]{
    font=footnotesize,
    skip=2pt
  }
  \captionsetup{
    skip=4pt
  }

  \subfloat[
    Mock 000:
    $0.0823\le z\le1.70$;
    $\langle\sigma_H\rangle=26.4$
    \label{fig:mock-0}
  ]{%
    \includegraphics[
      width=0.49\linewidth,
      trim=0 5mm 0 4mm,
      clip
    ]{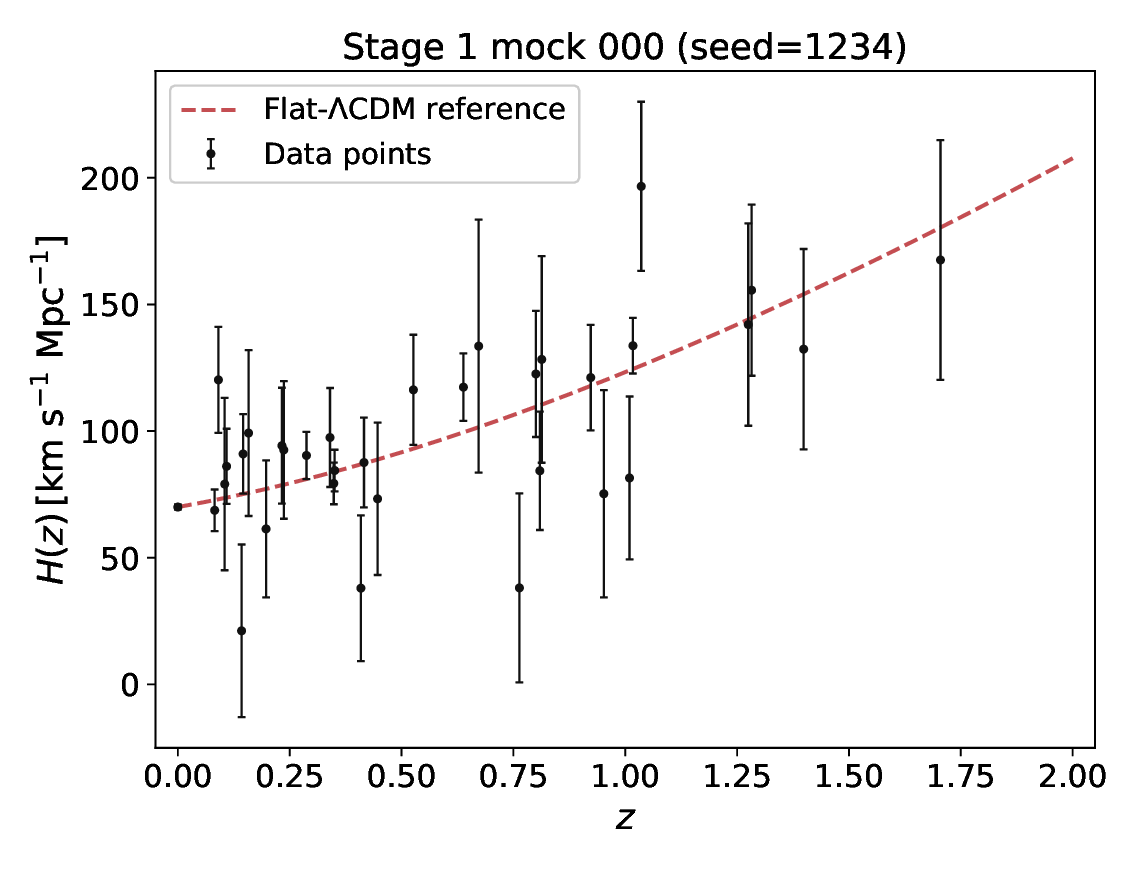}%
  }\hfill
  \subfloat[
    Mock 001:
    $0.0700\le z\le1.76$;
    $\langle\sigma_H\rangle=21.7$
    \label{fig:mock-1}
  ]{%
    \includegraphics[
      width=0.49\linewidth,
      trim=0 5mm 0 4mm,
      clip
    ]{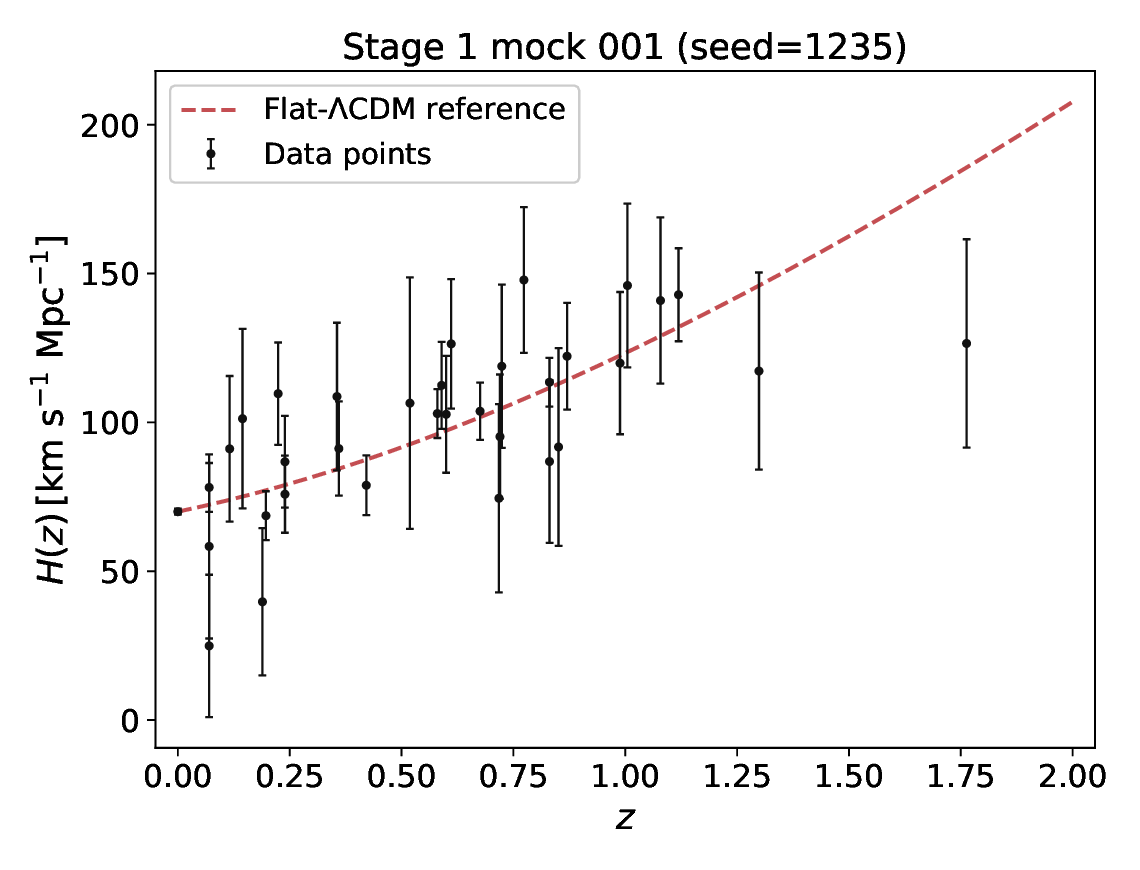}%
  }

  \par\vspace{2pt}

  \subfloat[
    Mock 002:
    $0.0700\le z\le1.97$;
    $\langle\sigma_H\rangle=22.4$
    \label{fig:mock-2}
  ]{%
    \includegraphics[
      width=0.49\linewidth,
      trim=0 5mm 0 4mm,
      clip
    ]{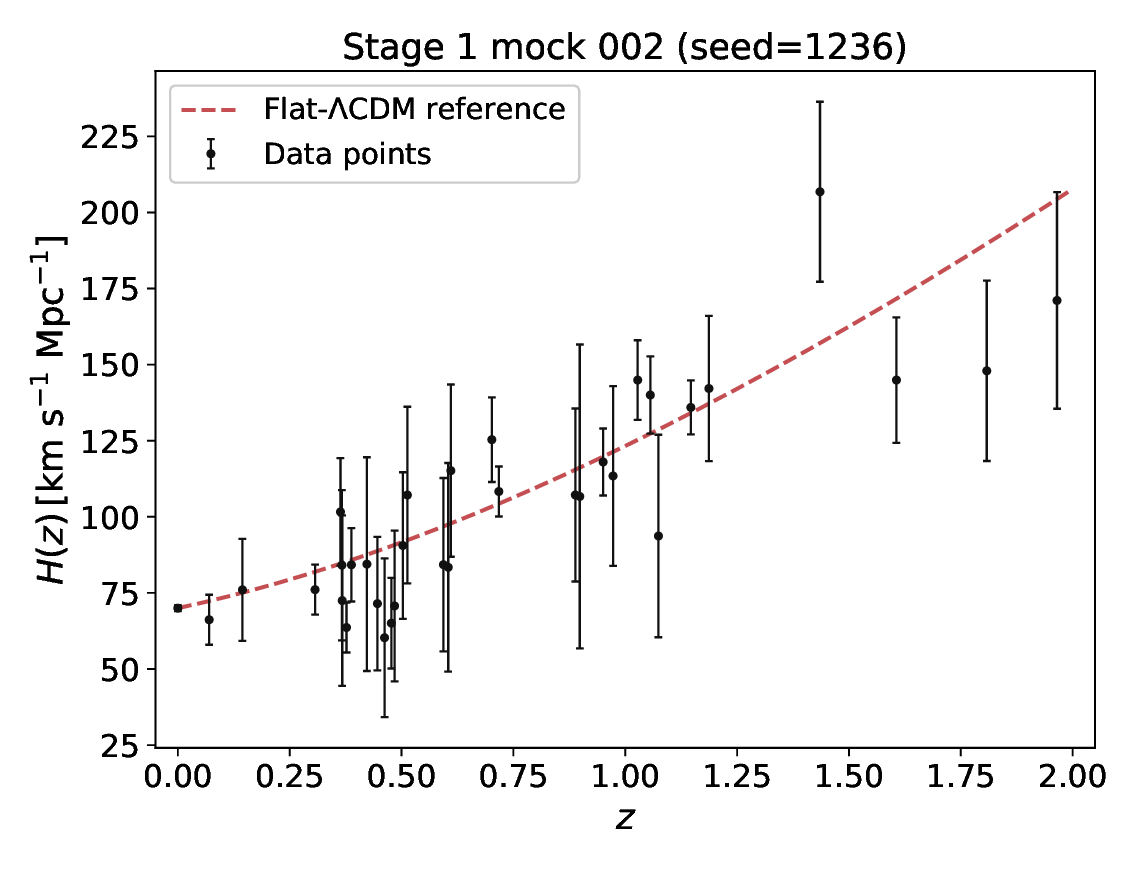}%
  }\hfill
  \subfloat[
    Mock 003:
    $0.0989\le z\le1.97$;
    $\langle\sigma_H\rangle=20.7$
    \label{fig:mock-3}
  ]{%
    \includegraphics[
      width=0.49\linewidth,
      trim=0 5mm 0 4mm,
      clip
    ]{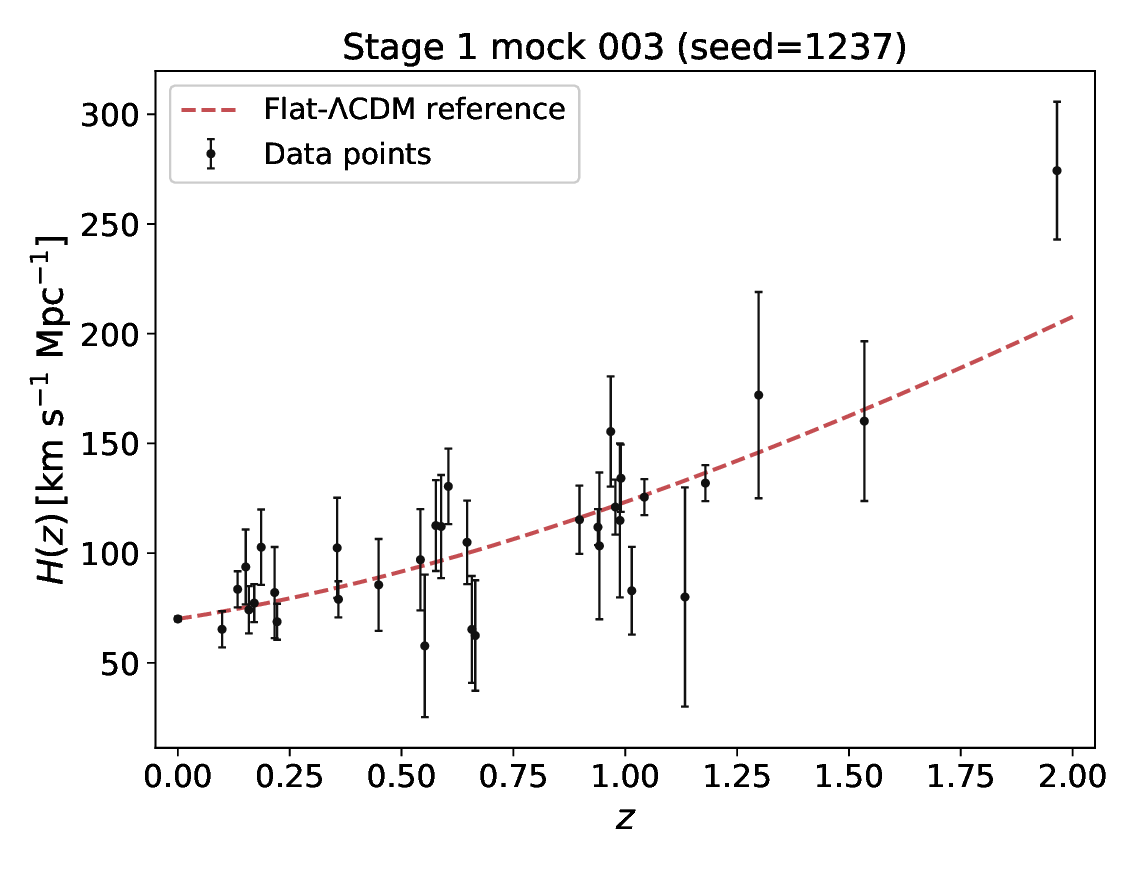}%
  }

  \par\vspace{2pt}

  \subfloat[
    Mock 004:
    $0.0850\le z\le1.97$;
    $\langle\sigma_H\rangle=23.2$
    \label{fig:mock-4}
  ]{%
    \includegraphics[
      width=0.49\linewidth,
      trim=0 5mm 0 4mm,
      clip
    ]{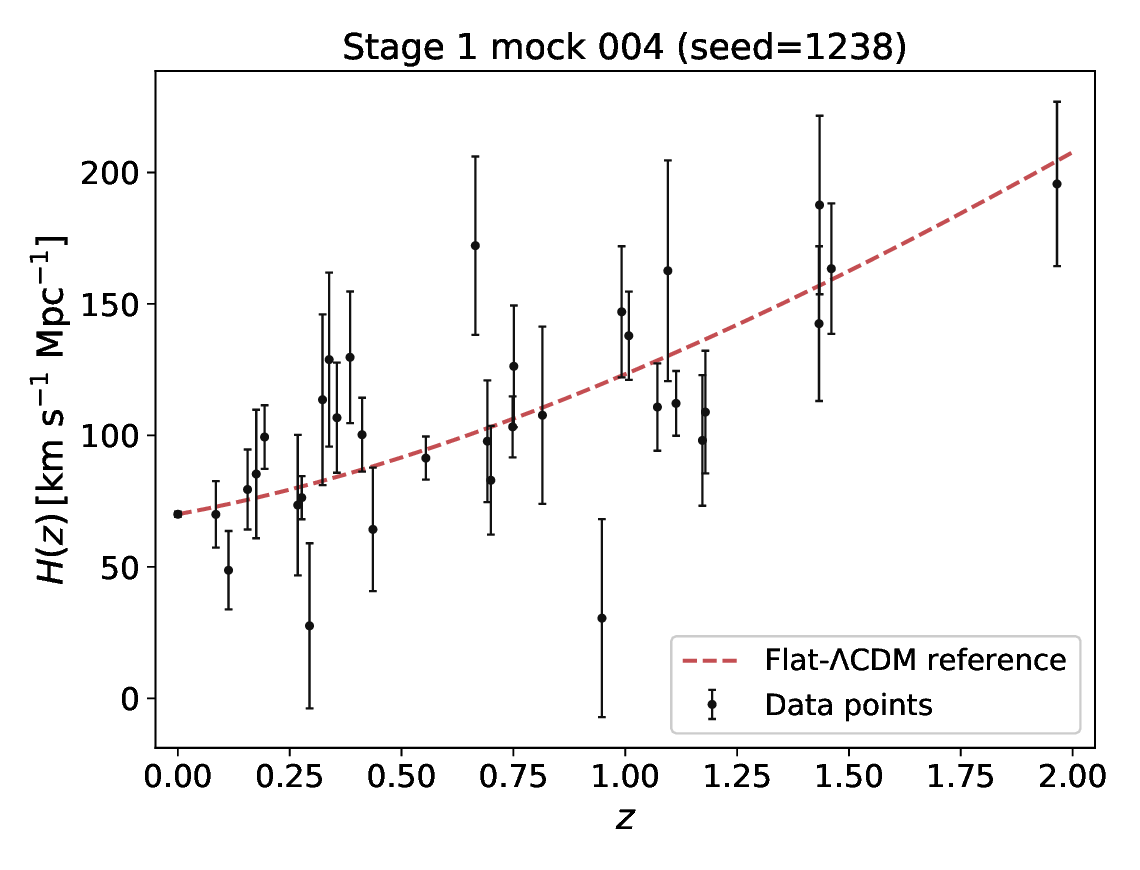}%
  }

  \caption{\label{fig:mock_details}
    The complete suite of five Stage-1 CC-like mock datasets used for every candidate in the Stage-2 full-factorial search. Black markers and vertical bars show $H^{\mathrm{mock}}(z)$ and its assigned $1\sigma$ input uncertainty, while the red dashed curve is the generating flat-$\Lambda$CDM background of \zcref{eq:hfid_stage1,eq:fiducial_values_stage1}. Each panel contains the fixed $z=0$ anchor and 33 stochastic CC-like points constructed in \zcref{subsec:mock_generation_cc}. The subcaptions give the realized non-anchor
redshift interval and non-anchor mean $\sigma_H$; dimensional $H$ and $\sigma_H$ values are in $\mathrm{km\,s^{-1}\,Mpc^{-1}}$.
}
\end{figure*}

In this appendix, we provide additional details on the mock datasets used in Stage-2. The five mock datasets are generated using different random seeds and cover a range of redshift intervals and observational uncertainties. Each dataset consists of 33 stochastic CC-like points along with a fixed $z=0$ anchor, constructed to mimic the characteristics of real cosmic chronometer data.
\zcref{fig:mock_details} shows all five mock datasets used as Stage-1 inputs in the full-factorial design search. The panels in this figure show the generated data themselves and their known fiducial background.

\clearpage

\section{\label{append:score_of_stage2-design}Score of Stage-2 design search}

The full-factorial Stage-2 results are listed in \zcref{tab:stage2_full_factorial}. The table combines the activation comparison and architecture comparison into a single 48-candidate summary, so that every activation--architecture pair is evaluated under the same aggregation rules. Each row contains five mock datasets and 100 seeds per mock ($n=500$ successful runs). Lower values are preferred for all Score columns and for the component error diagnostics.

\begin{table*}[tbp]
\caption{\label{tab:stage2_full_factorial}
Complete Stage-2 full-factorial activation--architecture search. Here $\mathcal{J}=\operatorname{mean}_{m}[\operatorname{median}_{s}(\mathrm{Score}_{m,s})]$ is the primary selection statistic, ``Pool med.'' and ``Pool mean'' are the median and mean over all 500 run-level Scores, and $\overline{S}_{\rm select}$ and $\overline{R}_{\rm smooth}$ are the corresponding pooled component means. $\overline{\rm RMSE}$ is measured in $\mathrm{km\,s^{-1}\,Mpc^{-1}}$, and $\overline{C}_{68}$ is the empirical 68\% coverage. The dagger marks the primary $\mathcal{J}$ minimum, SiLU-$[8]$, and the double dagger marks the pooled-median minimum, SiLU-$[16]$. The detailed table retains six decimal places for the Score-based statistics so that candidates that coincide under the three-significant-digit figure convention remain distinguishable; selection is performed with the full-precision values.}
\begingroup
\setlength{\tabcolsep}{3pt}
\renewcommand{\arraystretch}{1.05}
\scriptsize
\begin{ruledtabular}
\begin{tabular}{llcccccccc}
Activation & Architecture & $n_{\rm runs}$ & $\mathcal{J}$ & Pool med. & Pool mean & $\overline{S}_{\rm select}$ & $\overline{R}_{\rm smooth}$ & $\overline{\rm RMSE}$ & $\overline{C}_{68}$ \\
\hline
ELU & $[8]$ & 500 & 0.057680 & 0.053272 & 0.068614 & 0.065935 & 0.053579 & 5.561 & 0.993 \\
ELU & $[16]$ & 500 & 0.062340 & 0.058108 & 0.069245 & 0.067150 & 0.041899 & 5.752 & 0.995 \\
ELU & $[32]$ & 500 & 0.059462 & 0.059178 & 0.065008 & 0.062786 & 0.044444 & 5.555 & 0.996 \\
ELU & $[64]$ & 500 & 0.064848 & 0.063104 & 0.065875 & 0.063344 & 0.050616 & 5.402 & 0.994 \\
ELU & $[128]$ & 500 & 0.088426 & 0.086684 & 0.090434 & 0.086646 & 0.075770 & 5.967 & 0.989 \\
ELU & $[256]$ & 500 & 0.120752 & 0.122807 & 0.129796 & 0.124361 & 0.108686 & 6.559 & 0.988 \\
ELU & $[8,8]$ & 500 & 0.089380 & 0.086833 & 0.096813 & 0.093212 & 0.072026 & 6.946 & 0.988 \\
ELU & $[16,16]$ & 500 & 0.081175 & 0.080802 & 0.083008 & 0.078410 & 0.091965 & 6.212 & 0.992 \\
ELU & $[32,32]$ & 500 & 0.102525 & 0.094263 & 0.102895 & 0.094881 & 0.160285 & 6.465 & 0.990 \\
ELU & $[64,64]$ & 500 & 0.128557 & 0.141447 & 0.137120 & 0.122557 & 0.291255 & 7.116 & 0.987 \\
ELU & $[128,128]$ & 500 & 0.214207 & 0.199819 & 0.228060 & 0.197543 & 0.610332 & 9.711 & 0.977 \\
ELU & $[256,256]$ & 500 & 0.248379 & 0.237620 & 0.269390 & 0.229377 & 0.800267 & 10.263 & 0.976 \\
\hline
SiLU & $[8]$ & 500 & $0.047289^{\dagger}$ & 0.052320 & 0.075279 & 0.070759 & 0.090389 & 5.651 & 0.990 \\
SiLU & $[16]$ & 500 & 0.048620 & $0.047335^{\ddagger}$ & 0.062718 & 0.059623 & 0.061893 & 5.078 & 0.994 \\
SiLU & $[32]$ & 500 & 0.075548 & 0.068593 & 0.075923 & 0.072535 & 0.067761 & 5.709 & 0.992 \\
SiLU & $[64]$ & 500 & 0.096119 & 0.089378 & 0.096793 & 0.092633 & 0.083198 & 6.562 & 0.989 \\
SiLU & $[128]$ & 500 & 0.114554 & 0.095394 & 0.115381 & 0.110186 & 0.103897 & 7.275 & 0.988 \\
SiLU & $[256]$ & 500 & 0.129448 & 0.112151 & 0.134789 & 0.127082 & 0.154148 & 8.070 & 0.983 \\
SiLU & $[8,8]$ & 500 & 0.091969 & 0.087230 & 0.108849 & 0.104010 & 0.096789 & 7.105 & 0.982 \\
SiLU & $[16,16]$ & 500 & 0.100103 & 0.088480 & 0.107498 & 0.101024 & 0.129494 & 7.211 & 0.987 \\
SiLU & $[32,32]$ & 500 & 0.146008 & 0.147972 & 0.140323 & 0.128768 & 0.231109 & 8.699 & 0.984 \\
SiLU & $[64,64]$ & 500 & 0.179142 & 0.186519 & 0.177937 & 0.160825 & 0.342240 & 10.002 & 0.978 \\
SiLU & $[128,128]$ & 500 & 0.199505 & 0.190318 & 0.204772 & 0.183618 & 0.423077 & 10.380 & 0.976 \\
SiLU & $[256,256]$ & 500 & 0.204133 & 0.204304 & 0.214943 & 0.182671 & 0.645439 & 10.202 & 0.974 \\
\hline
$\tanh$ & $[8]$ & 500 & 0.100407 & 0.097769 & 0.107170 & 0.102700 & 0.089406 & 7.758 & 0.985 \\
$\tanh$ & $[16]$ & 500 & 0.093120 & 0.091537 & 0.100653 & 0.096613 & 0.080797 & 7.442 & 0.989 \\
$\tanh$ & $[32]$ & 500 & 0.086483 & 0.087960 & 0.087546 & 0.083204 & 0.086840 & 6.856 & 0.994 \\
$\tanh$ & $[64]$ & 500 & 0.084385 & 0.080772 & 0.085521 & 0.080564 & 0.099146 & 6.502 & 0.993 \\
$\tanh$ & $[128]$ & 500 & 0.091865 & 0.076594 & 0.091991 & 0.084818 & 0.143463 & 6.290 & 0.990 \\
$\tanh$ & $[256]$ & 500 & 0.111503 & 0.109815 & 0.117645 & 0.109049 & 0.171924 & 6.361 & 0.988 \\
$\tanh$ & $[8,8]$ & 500 & 0.113396 & 0.112225 & 0.124901 & 0.116544 & 0.167145 & 8.451 & 0.979 \\
$\tanh$ & $[16,16]$ & 500 & 0.106334 & 0.111867 & 0.110927 & 0.101533 & 0.187886 & 7.733 & 0.986 \\
$\tanh$ & $[32,32]$ & 500 & 0.122878 & 0.124752 & 0.125880 & 0.110627 & 0.305050 & 7.511 & 0.983 \\
$\tanh$ & $[64,64]$ & 500 & 0.157591 & 0.123217 & 0.164050 & 0.137993 & 0.521137 & 7.766 & 0.982 \\
$\tanh$ & $[128,128]$ & 500 & 0.206861 & 0.185496 & 0.219694 & 0.180128 & 0.791322 & 8.805 & 0.980 \\
$\tanh$ & $[256,256]$ & 500 & 0.294502 & 0.279554 & 0.310688 & 0.229357 & 1.626613 & 10.319 & 0.975 \\
\hline
TanhExp & $[8]$ & 500 & 0.067702 & 0.066979 & 0.088130 & 0.083710 & 0.088393 & 6.325 & 0.987 \\
TanhExp & $[16]$ & 500 & 0.072326 & 0.070843 & 0.081665 & 0.077627 & 0.080748 & 6.179 & 0.991 \\
TanhExp & $[32]$ & 500 & 0.088494 & 0.078468 & 0.087733 & 0.082707 & 0.100509 & 6.309 & 0.989 \\
TanhExp & $[64]$ & 500 & 0.096376 & 0.079730 & 0.097752 & 0.091626 & 0.122524 & 6.544 & 0.988 \\
TanhExp & $[128]$ & 500 & 0.104152 & 0.084127 & 0.105498 & 0.097064 & 0.168673 & 6.863 & 0.988 \\
TanhExp & $[256]$ & 500 & 0.132141 & 0.128362 & 0.137998 & 0.125884 & 0.242263 & 8.014 & 0.983 \\
TanhExp & $[8,8]$ & 500 & 0.098258 & 0.090525 & 0.114851 & 0.106899 & 0.159052 & 7.551 & 0.983 \\
TanhExp & $[16,16]$ & 500 & 0.096429 & 0.095606 & 0.101005 & 0.090907 & 0.201948 & 6.707 & 0.988 \\
TanhExp & $[32,32]$ & 500 & 0.145325 & 0.143780 & 0.147874 & 0.130927 & 0.338937 & 8.736 & 0.980 \\
TanhExp & $[64,64]$ & 500 & 0.197688 & 0.190236 & 0.201749 & 0.176884 & 0.497293 & 10.327 & 0.976 \\
TanhExp & $[128,128]$ & 500 & 0.229639 & 0.232115 & 0.233689 & 0.195398 & 0.765815 & 10.370 & 0.974 \\
TanhExp & $[256,256]$ & 500 & 0.292916 & 0.321611 & 0.299676 & 0.186535 & 2.262809 & 9.875 & 0.974 \\
\end{tabular}
\end{ruledtabular}
\endgroup
\end{table*}
\clearpage

\bibliographystyle{apsrev4-2}
\bibliography{refs}

\end{document}